\pdfoutput=1
\documentclass[12pt,a4paper]{article}
\usepackage{a4wide}

\usepackage{subcaption}
\usepackage[english]{babel}
\usepackage{amsmath}   
\usepackage{graphicx}  
\usepackage{bm}        
\usepackage{amssymb}   
\usepackage{epsfig}
\usepackage{epstopdf}
\usepackage{amsmath}
\usepackage{fontenc}
\usepackage[toc]{appendix}
\usepackage{color,soul} 
\usepackage{datetime}
\usepackage{physics}
\usepackage{enumerate}
\usepackage{multirow}

\usepackage{footmisc}
\usepackage{textpos}
\usepackage{caption}
\usepackage{boxhandler}

\usepackage{hyperref}

\newcommand{\be}{\begin{equation}}
\newcommand{\ee}{\end{equation}}
\newcommand{\bea}{\begin{eqnarray}}
\newcommand{\eea}{\end{eqnarray}}
\newcommand{\bean}{\begin{eqnarray*}}
\newcommand{\eean}{\end{eqnarray*}}
\newcommand{\nn}{\nonumber}

\newcommand{\cA}{\mathcal{A}}

\newcommand{\cO}{\mathcal{O}}

\newcommand{\cE}{\mathcal{E}}

\begin{document}



\vspace*{-14mm}

\begin{center}
\begin{center}
{\fontsize{20.74}{20.74} \bf{On Detecting Equatorial Symmetry Breaking with LISA}} \medskip \\
\end{center}
\vspace{2mm}

\centerline{{\bf Kwinten Fransen, Daniel R. Mayerson}}
\vspace{8mm}

\centerline{Institute for Theoretical Physics, KU Leuven,
            Celestijnenlaan 200D, B-3001 Leuven, Belgium}
\vspace{6mm}

{\footnotesize\upshape\ttfamily kwinten.fransen, daniel.mayerson @ kuleuven.be} \\

\vspace{6mm}
 
\textsc{Abstract}

\end{center}

\noindent The equatorial symmetry of the Kerr black hole is generically broken in models of quantum gravity. Nevertheless, most phenomenological models start from the assumption of equatorial symmetry, and little attention has been given to the observability of this smoking gun signature of beyond-GR physics. Extreme mass-ratio inspirals (EMRIs), in particular, are known to sensitively probe supermassive black holes near their horizon; yet estimates for constraints on deviations from Kerr in space-based gravitational wave observations (e.g. with LISA) of such systems are currently based on equatorially symmetric models.
We use modified ``analytic kludge'' waveforms to estimate how accurately LISA will be able to measure or constrain equatorial symmetry breaking, in the form of the lowest-lying odd parity multipole moments $S_2, M_3$. We find that the dimensionless multipole ratios such as $S_2/M^3$ will typically be detectable for LISA EMRIs with a measurement accuracy of $\Delta(S_2/M^3) \sim 1\%$; this will set a strong constraint on the breaking of equatorial symmetry.

\vspace{-4mm}

\thispagestyle{empty}

\setcounter{tocdepth}{2}
\tableofcontents

\newpage
\section{Introduction}\label{sec:intro}

Black holes will be probed to high precision by gravitational wave astronomy in the coming decades. The inspiral and capture of stellar mass compact objects into supermassive black holes holds particular promise \cite{berry2019astro2020}. From such extreme mass-ratio inspiral (EMRI) events, it is estimated that LISA, a space-based observatory \cite{baker2019laser}, could determine the mass and spin of the supermassive black hole to about one part in $10^5$ \cite{Barack:2003fp,Babak:2017tow,Gair:2017ynp}. 

A remarkable prediction of general relativity is that the mass and spin of the black hole are its only distinguishing properties (in vacuum). EMRIs will be a powerful tool to search for observational evidence to the contrary. In particular, all of the non-zero multipoles of Kerr are determined by its mass $M_0=M$ and spin $S_1=J=Ma$:
\be M_{2\ell} = M(-a^2)^\ell, \qquad S_{2\ell+1} = Ma(-a^2)^\ell.\ee
The first multipole for which the Kerr solution then gives a non-trivial prediction is the mass quadrupole moment $M_2 = -S_1^2/M$. LISA will be able to measure this dimensionless multipole ($M_2/M^3$) below the 1\%-level \cite{Babak:2017tow,Gair:2017ynp,Barack:2006pq}. 

The odd-parity multipoles $M_{2\ell+1}, S_{2\ell}$ vanish identically for Kerr, implying it is equatorially symmetric: the metric remains invariant when reflected over the equatorial plane. 
This equatorial symmetry of Kerr is ``accidental'', in the sense that there is no underlying reason for its existence; this is in contrast to axisymmetry, which is a consequence of stationarity for vacuum black holes in GR \cite{Townsend:1997ku}. As such, there is no reason for equatorial symmetry \emph{not} to be broken in beyond-GR physics. Indeed, equatorial symmetry \emph{is} generically broken in many models such as (odd-parity) higher-derivative corrections to GR \cite{Endlich:2017tqa,Cardoso:2018ptl,Cano:2019ore}, string theory black holes \cite{Bena:2020uup,Bah:2021jno}, and compact, horizonless objects such as fuzzballs \cite{Bena:2020see,Bianchi:2020bxa,Bena:2020uup,Bianchi:2020miz,Mayerson:2020tpn,Bah:2021jno}. 

Even though equatorial symmetry breaking is ubiquitous in beyond-GR physics, many gravitational phenomenology (including EMRI investigations) either assume an equatorial reflection symmetry and explicitly set $M_{2\ell+1}=S_{2\ell}=0$ \cite{Ryan:1995wh, Glampedakis:2005cf, Gair:2007kr,Moore:2017lxy}, or restrict to $M_2$ deformations \cite{Barack:2006pq, Babak:2017tow,Gair:2017ynp}. $S_2$ and $M_3$ are, a priori, the next most important multipoles. However, they are also the first multipoles that break equatorial symmetry, and are therefore of a qualitatively different nature; it is not obvious how well results based on equatorially symmetric multipoles should generalize.   We are thus left with the burning question: how well will we be able to detect this phenomenon with EMRI observations? 

There are a few studies that have considered some aspects of equatorial symmetry breaking (although sometimes very briefly). Various aspects of equatorial symmetry-breaking spacetimes  are discussed in \cite{Raposo:2018xkf,Cunha:2018uzc,Cardoso:2018ptl,Datta:2020axm,Lima:2021las}, but none of these include a detailed analyses on its measurability.
Multipole moments of the equatorial symmetry-breaking Kerr-NUT spacetime were considered in \cite{Mukherjee:2020how}; this included an analyses of how its multipoles affect the orbital frequencies in gravitational wave signals for near-circular, near-equatorial orbits (generalizing Ryan \cite{Ryan:1995wh} and similar to our Section \ref{sec:circular} below) --- note that Kerr-NUT breaks asymptotic flatness as $S_0\neq 0$.
Notably, for asymptotically flat spacetimes, \cite{Sopuerta:2009iy} discusses the (significant) potential of EMRIs to constrain dynamical Chern-Simons theory, which modifies the Kerr multipoles starting from $S_4$. On the other hand, \cite{Endlich:2017tqa} discusses the gravitational radiation effects due to (other) higher-derivative corrections (including odd-parity ones), and estimates that current measurements cannot constrain these parameters much. 
Finally, odd-parity multipoles featured briefly in ``bumpy'' black hole analyses \cite{Vigeland:2010xe, Vigeland:2009pr}, where it was found that there was no average influence of odd-parity bumps on orbital frequencies. These results seem discouraging for the measurability of equatorial symmetry breaking --- although we will show here that a pessimistic conclusion would be too rash.

We will use the ``analytic kludge'' formalism developed by Barack and Cutler \cite{Barack:2003fp,Barack:2006pq} to investigate the accuracy that LISA can measure the equatorial symmetry-breaking dimensionless multipoles $S_2/M^3$ and $M_3/M^4$ for EMRIs. We will find that these generally can be measured and constrained to within $\sim 10^{-2}$. LISA measurement of EMRIs will thus give a surprisingly accurate and stringent measurement and contraint on the breaking of equatorial symmetry, a smoking gun of beyond-GR physics!

\medskip

Before turning to general EMRIs, we first discuss the less general case of near-circular, near-equatorial orbits in Section \ref{sec:circular}. This was originally investigated by Ryan \cite{Ryan:1995wh}, and an initial analysis of the measurement accuracies of various (even-parity) multipoles for such orbits was performed also by Ryan \cite{Ryan:1997hg}.

We then introduce the analytic kludge formalism for generating EMRI waveforms in Section \ref{sec:generic} and discuss how to generalize it to include the effects of the equatorial symmetry-breaking multipoles $S_2,M_3$. Our main results for LISA parameter estimation accuracy, based on the analytic kludge and a Fisher analysis, are presented in Section \ref{sec:measure}. We show that $S_2/M^3$ can be measured to within $\sim 10^{-2}$, and discuss the dependence of this prediction on the orbital parameters. Finally, in Section \ref{sec:discussion}, we discuss our results and their implications.
\section{Near-Circular, Near-Equatorial Orbits}\label{sec:circular}

Throughout we will consider the gravitational two-body problem where the two bodies have masses $\mu,M$ with $\mu\ll M$. These are called extreme mass-ratio inspirals (EMRIs). The supermassive object will have a nontrivial multipolar structure which we would like to constrain with gravitational wave observations, the stellar mass object will be modeled as a featureless point particle\footnote{We use geometric units $G=1$ and $c=1$.}.

Typical EMRIs are expected to have a rich and interesting orbital evolution, which in particular will be eccentric and inclined \cite{Gair:2017ynp}. However, before dealing with this general case, in this Section we consider the adiabatic evolution of near-equatorial circular orbits. When odd-parity, equatorial symmetry-breaking multipoles such as $S_2$ and $M_3$ are non-zero, a purely equatorial orbit is not possible \cite{Datta:2020axm}. 
 In this case, near-equatorial circular orbits are possible with the relative separation vector\footnote{The coordinates used are those in \eqref{eqn:metric}.}
\begin{equation}\label{eqn:circularseparation}
	\vec{r} = r \sin \xi  \begin{pmatrix}
		\cos{(2 \pi \nu t + \frac{\pi}{2})}\, , & \sin{(2 \pi \nu t + \frac{\pi}{2})} \, , & \cot\xi
	\end{pmatrix} \, .
\end{equation}
Here, $r$ is the radial separation, $\nu$ is the orbital frequency and $\xi$ is the relative inclination with respect to the direction of the orbital angular momentum. Explicitly, in a large separation expansion in terms of the fiducial relative velocity $v=(2 \pi \nu M)^{1/3}$, this inclination is given by  
\begin{align} \label{eqn:xi}
	\cos{\xi} &= -\frac{3 S_2}{M^3}v^5 -\frac{3 M_3}{2 M^4}v^6 -\frac{4 S_2}{M^3} v^7 +v^8 \left(- \frac{9M_3}{2 M^4} - \frac{21 S_1 S_2}{M^5} \right) \\
	&- v^9\left(\frac{12 M_3 S_1}{M^6}+\frac{15 M_2 S_2}{M^6}+\frac{101 S_2}{14
		M^3}-\frac{15 S_4}{4 M^5}\right) + \ldots \, .\nn
\end{align}
On the other hand, the equatorially asymmetric multipoles responsible for this inclination contribute only quadratically to the binding energy $E$ and separation $r$ in function of the orbital frequency\footnote{Note that we present this particular form of the frequency derivative of $E$ in order to allow for easy comparison with \cite{Ryan:1995wh}.}
\begin{align}
	\delta r_{\text{asym}}  &= \frac{19}{2} M v^8  \frac{S_2^2}{M^6} + 12 M v^9 \frac{S_2 M_3}{M^7} + M v^{10} \left( \frac{15}{4} \frac{M_3^2}{M^8} + \frac{23789}{792} \frac{S_2^2}{M^6} \right) + \ldots	\, , \\
	\delta( -\frac{\nu}{\mu} \frac{d E}{d \nu})_{\text{asym}} &= -57 v^{12} \frac{S^2_2}{M^6} - 65 v^{13} \frac{M_3 S_2}{M^7} + v^{14} \left( -\frac{77}{4} \frac{M_3^2}{M^8} - \frac{18577}{54} \frac{S_2^2}{M^6} \right) + \ldots \, .
\end{align}
The contributions by the other, equatorially symmetric, multipoles are well-known \cite{Ryan:1995wh}. We give the full expressions up to respectively $\cO(v^{10})$ and  $\cO(v^{14})$ in \eqref{eqn:fullcircr} and \eqref{eqn:fullDeltaE}  in appendix \ref{app:circular}, where we also provide more details on the derivation of these results.

The quadrupole formula provides the leading order radiation reaction\footnote{Except for the contribution from the spin $S_1$, where one also needs to take into account the current quadrupole emission.}, which can again be seen to be corrected only to quadratic order by $S_2,M_3$:
\begin{equation}
	- \frac{d E}{d t} = \frac{32 \mu^2 r^4 \sin^2 \xi  (2 \pi \nu)^6}{5}(1+\cos^2 \xi) \, .\label{eqn:ppquadrupole}
\end{equation}
Note that the emission pattern and frequency content are already modified at linear order and behave as what would ordinarily be a current quadrupole emission. 
This phenomenon occurs also from parity violating interactions \cite{Endlich:2017tqa}. It would be interesting to investigate the signature of such emission, for instance in the case of comparable mass binaries where it is otherwise dynamically suppressed. Nevertheless, our focus will remain on the likely scenario that only the dominant gravitational wave emission is observed. For us, the influence of (unexpected) multipole moments is then due to its modification of the orbital dynamics.

From the orbital and radiated energies,  \eqref{eqn:fullDeltaE} and \eqref{eqn:ppquadrupole}, as a function of the frequency, one can derive, in an adiabatic approximation, the correction of the multipoles to the gravitational waveform. We will additionally use a stationary phase approximation to go to the frequency domain. Finally, we will focus on the change in phase such that the resulting frequency-domain waveform has the following structure
\begin{equation}\label{eqn:fdomainwave}
	\tilde{h}(f) = \cA f^{-7/6} e^{i \psi(f)} \, , \quad  \psi(f)= \psi_0(f)+ \delta \psi(f) \, ,
\end{equation} 
with $f = 2\nu$, the gravitational wave frequency. Here, $\psi_0(f)$ represents the point-particle, or non-multipolar contribution. We will simply approximate it with a 3.5PN TaylorF2 phasing \cite{Buonanno:2009zt}. Many other choices could have been made, but this will not make a significant difference for our purposes. In particular, we have verified this by comparing with the choice of \cite{Ryan:1997hg} which essentially amounts to a 4PN, adiabatic extreme mass-ratio inspiral \cite{tagoshi1994post, Poisson:1994yf, Poisson:1995vs, Tanaka:1996lfd}. Instead, $\delta \psi(f)$ is the leading contribution of the multipoles. Of particular interest here is the leading equatorial symmetry breaking contribution
\begin{equation}
	\delta \psi_{\text{asym}}(f) = \frac{3}{128}(\frac{M}{\mu})(\pi M f)^{5/3}\left( 908  (\frac{S_2}{M^3})^2  - 580 (\pi M f)^{1/3} \frac{S_2 M_3}{M^7}  - \frac{1545}{14}(\pi M f)^{2/3} (\frac{M_3}{M^4})^2  \right) \, .
	\label{eqn:oddpsi}
\end{equation}
The full $\delta \psi(f) = \delta \psi_{\text{sym}}+\delta \psi_{\text{asym}}$ is then found by including also $\delta \psi_{\text{sym}}$ which was previously derived \cite{Ryan:1997hg} and is reproduced here as \eqref{eqn:evenpsi} in the appendix.

In order to provide an initial estimate of how well multipoles can be measured given this waveform model, we use a Fisher matrix analysis assuming a stationary, Gaussian noise. This means that for a choice of free parameters $\vec{\theta} = \begin{pmatrix}
	t_* \, , \phi_* \, , \mu \, , M \, , S_1 \, , M_2 \, , S_2^2\, , \ldots,
\end{pmatrix}$, where $t_*$, $\phi_*$ are a reference time and phase, we compute
\begin{equation}\label{eqn:fisher}
	(\Gamma)_{ij}=\left\langle \frac{\partial h}{\partial \theta^{i}} , \frac{\partial h}{\partial \theta^{j}} \right \rangle \, , \quad \text{with} \quad \left\langle h_1 , h_2 \right \rangle = 2 \int_0^{\infty} df \frac{\tilde{h}^*_1(f) \tilde{h}_2(f)+\tilde{h}_1(f) \tilde{h}^*_2(f)}{S_n(f)}\, ,
\end{equation}
to find the covariance matrix $\sigma = \Gamma^{-1}$. The measurement accuracies are then approximated by the standard deviations $\Delta \theta^{i} = \sqrt{(\Sigma)_{ii}}$. We use the LISA noise curve from \cite{Robson:2018ifk} for the noise spectral density $S_n(f)$, see \eqref{eqn:PSD} for an explicit expression. Importantly, at this stage we do not yet take into account the movement of LISA and consequently the amplitude $\mathcal{A}$ in \eqref{eqn:fdomainwave} is assumed to be constant, and set by a choice of signal-to-noise (SNR).

	\begin{table}[ht]\centering 
		 \begin{tabular}{ c c c c c c c c c   } 
			\hline
			\hline
			\rule{0pt}{3ex} $L(\Delta t_*)$ & $L(\Delta \phi_*)$ & $L(\frac{\Delta \mu}{\mu})$ & $L(\frac{\Delta M}{M})$ & $L(\Delta \tilde{S}_1$) & $L(\Delta \tilde{M}_2$) & $L(\mathbf{\Delta \tilde{S}^2_2})$ & $L(\mathbf{\Delta \tilde{S}_2 \tilde{M}_3})$ & $L(\mathbf{\Delta \tilde{M}^2_3})$  \\ [0.5ex] 
			\hline
			$0.8$ & $-0.2$ & $-5.1$ & $-5.5$ & & & & &  \\ 
			
			$2.3$& $2.2$ & $-3.1$ & $-3.3$ & $-2.9$ &  & & & \\
			
			$2.5$& $2.3$ & $-2.9$ & $-3.1$ & $-2.8$ & $-3.2$ &  & &   \\
			
			$3.3$ & $2.8$ & $-2.4$ & $-2.6$ &  $-2.3$ & $-3.0$ & $\mathbf{-0.8}$  &  & \\
			$5.7$ & $4.7$ & $-0.3$ & $-0.6$ &  $-0.8$ & $0.4$ & $\mathbf{1.8}$  & $\mathbf{2.2}$ & $\mathbf{3.4}$ \\
			\hline
			\hline
			
		\end{tabular}
		\caption{The errors for the different parameters in the waveform model \eqref{eqn:fdomainwave} when including more and more multipole moments given 1 year of LISA observation before the ISCO for SNR=30, assuming the multipoles vanish, $M=10^5 M_{\odot}$ and $\mu = 10M_{\odot}$. We abbreviate $\log_{10}(\ldots)=L(\ldots)$ and $\tilde{S}_l = \frac{S_l}{M^{l+1}}$, $\tilde{M}_l = \frac{M_l}{M^{l+1}}$.}
		\label{tbl:ryan}
	\end{table}
	
In Table \ref{tbl:ryan}, the results of this analysis are shown as more and more multipoles are included as free parameters, up to $M^2_3$, at which point there are already too many parameters to meaningfully constrain each additional multipole individually. As a check, we have ensured that, without the novel corrections and with a matching noise spectral density ($S_n$ in \eqref{eqn:fisher}), we reproduce the results of \cite{Ryan:1997hg}.

Although the number of new parameters quickly proliferates and undermines the determination of individual multipoles in this approach, Table \ref{tbl:ryan} still suggests that $S_2$ could be measurable to a reasonable accuracy. To support this, the same analysis has been performed but only adding individual extra multipoles as free parameters, one at a time. Such an analysis yields, for instance, $\Delta (S^2_2/M^6) \sim 10^{-1}$ as well as $\Delta (M^2_3/M^8) \sim 10^{-1}$ given $\mu = 10M_{\odot}$, $M=10^5 M_{\odot}$ at an SNR of 30. For comparison, it gives $\Delta (M_2/M^3) \sim 10^{-3}$. An extended table of these results can be found in Table \ref{tbl:ryanindividually} in appendix \ref{app:circular}. The degradation of measurement accuracies in Table \ref{tbl:ryan} then indeed largely follows from the increasing number of parameters. 

To summarize, odd-parity multipoles, breaking equatorial symmetry, are qualitatively different from their even-parity counterparts. It is therefore unlikely that the present understanding of EMRIs as superb probes of multipolar structures can simply be extrapolated to include them. 
On the contrary, they do not affect the gravitational wave signal to linear order for near-circular near-equatorial orbits. Therefore, the ability of LISA to constrain them could be a lot \emph{worse}. However, as we show in the following Sections, this does not turn out to be the case when we provide a more realistic estimate of LISA's potential to constrain equatorial symmetry breaking in the form of these odd parity multipoles.

\section{Generic Orbits: The Analytic Kludge}\label{sec:generic}

To get a better idea of how precisely equatorial symmetry breaking would be measurable, we want to move away from near-equatorial, near-circular orbits as well as model LISA's detections more realistically. We need a way to simulate inspiral waveforms for general EMRI's and investigate what the effects on these waveforms are when non-zero odd-parity multipoles $S_2,M_3$ are present. Generating accurate waveforms for generic (Kerr) orbits is a difficult problem \cite{Poisson:2011nh, Barack:2018yvs, Pound:2021qin}, that has so far been solved to adiabatic order \cite{Hughes:2021exa} although with additionally an understanding of the full first order self-force \cite{vandeMeent:2017bcc}. It is an active area of research both to go beyond adiabatic order \cite{Warburton:2021kwk, Wardell:2021fyy} as well as to improve computational efficiency \cite{VanDeMeent:2018cgn, Chua:2020stf, Katz:2021yft}.

 In view of this, various approximate methods have been developed. We will use as our starting point the ``analytic kludge'' waveforms of Barack and Cutler \cite{Barack:2003fp}. These have the advantage of being simpler to compute than other, more accurate waveforms such as the ``numerical kludge'' methods \cite{Glampedakis:2002cb, Glampedakis:2002ya,Gair:2005ih,Babak:2006uv}, while also being easier to adapt to (unknown) non-Kerr spacetimes than say ``augmented analytic kludges''\cite{Chua:2015mua, Chua:2017ujo}\footnote{Although see e.g. \cite{Liu:2020ghq} for an adaption with a differing quadrupole moment.} or ``effective-one-body'' models \cite{Yunes:2009ef, Yunes:2010zj, Albanesi:2021rby}. The analytic kludge waveforms were used by Barack and Cutler to estimate that LISA could measure the masses of both EMRI bodies as well as the massive BH spin to fractional accuracy $\sim 10^{-5}-10^{-4}$ \cite{Barack:2003fp}, and the quadrupole $M_2$ to within $\sim 10^{-4}-10^{-2}$ \cite{Barack:2006pq}. These numbers seem to be robust, despite the shortcomings of the model \cite{berry2019astro2020,Huerta:2008gb,Liu:2020ghq}. Therefore, although the use of the analytic kludge is a sacrifice in waveform accuracy, it is more than sufficient for our initial, proof of principle analysis of measuring equatorial symmetry breaking. 

\subsection{Setup and parameter space}

In the analytic kludge, the EMRI is approximated as an instantaneous Newtonian-orbit binary which emits a quadrupolar waveform. Post-Newtonian equations are used to secularly evolve the orbit parameters. The approximated EMRI orbit is then translated into an observed waveform, taking into account the motion of the LISA detector using a low-frequency approximation \cite{Cutler:1997ta}. (For more details, see \cite{Barack:2003fp,Barack:2006pq}.)

A binary system where both objects are Kerr black holes would be described by 17 parameters. However, we will follow \cite{Barack:2003fp,Barack:2006pq} in neglecting the smaller object's spin, reducing the number of parameters to 14. We then add three additional parameters to allow for the possibility that the multipoles $M_2,S_2,M_3$ can differ from the Kerr values $M_2=-S_1^2/M$ and $S_2=M_3=0$. We are left with 17 parameters, summarized in Table \ref{tab:paramsummary}.

\begin{table}[ht]
\begin{tabular}{c|c|l}
\hline\hline
$\lambda^0$ & $t_0(\times 1\text{mHz})$ & total inspiral orbit time\\
$\lambda^1$ & $\ln \mu$ & (log of) smaller object's mass\\
$\lambda^2$ & $\ln M$ & (log of) large, central BH's mass\\
$\lambda^3$ & $\tilde S_1 = S_1/M^2$ & large BH's dimensionless spin magnitude\\
$\lambda^4$ & $e_0$ & final value (i.e. at LSO) of orbit eccentricity\\
$\lambda^5$ & $\tilde\gamma_0$ & final value  for $\tilde\gamma$ (the angle between $\hat{L}\times\hat{S}$ and pericenter)\\
$\lambda^6$ & $\Phi_0$ & final value for mean anomaly $\Phi$\\
$\lambda^7$ & $\mu_S:=\cos\theta_S$ & (cosine of) source's direction's polar angle\\
$\lambda^8$ & $\phi_S$ & azimuthal direction to source\\
$\lambda^9$ & $\cos\lambda$ & (cosine of) orbit inclination angle ($\hat{L}\cdot \hat{S}$)\\
$\lambda^{10}$ & $\alpha_0$ & final value of azimuthal angle $\alpha$ of $\hat{L}$ in the orbital plane\\
$\lambda^{11}$ & $\mu_K := \cos\theta_K$ & (cosine of) polar angle of large BH spin\\
$\lambda^{12}$ & $\phi_K$ & azimuthal direction of large BH spin\\
$\lambda^{13}$ & $\ln(\mu/D)$ & (log of) smaller object's mass divided by distance to source\\
$\lambda^{14}$ & $\tilde M_2 = M_2/M^3+\tilde S_1^2$ & large BH's dimensionless (mass) quadrupole moment\\
$\lambda^{15}$ & $\tilde S_2 = S_2/M^3$ & large BH's dimensionless current quadrupole moment\\
$\lambda^{16}$ & $\tilde M_3 = M_3/M^4$ & large BH's dimensionless mass octupole moment\\
\hline\hline
\end{tabular}
\caption{Summary of the 17 parameters of the EMRI inspiral. Note that when the large central black hole is a Kerr BH, it satisfies $\tilde M_2=\tilde S_2=\tilde M_3=0$ and $0\leq \tilde S_1\leq 1$.}
\label{tab:paramsummary}
\end{table}

The parameter $t_0$ indicates the time at which the smaller object reaches its last stable orbit (LSO), where we end the integration and the inspiral transitions to a plunge. The masses of the smaller and large object are respectively $\mu$ and $M$.

The angles $(\theta_S,\phi_S,\theta_K,\phi_K)$ specify the orientation of the orbit and central black hole spin with respect to an ecliptic-based coordinate system. The distance to the source is given by $D$.

The parameters $e,\tilde\gamma,\alpha,\lambda,\Phi$ correspond to orbital elements for the smaller object's trajectory, as defined relatively to the central black hole's spin (i.e. taking the unit vector $\hat{S}$ to lie along the positive $z$-axis). The orbit eccentricity is $e$. The angle $\tilde\gamma$ is the angle (in the plane of the orbit) from $\hat{L}\times \hat{S}$ to the pericenter (where $\hat{L}$ is the unit vector of the orbit's angular momentum) --- in standard orbit element terminology, this would correspond to the argument of periapsis, i.e. the angle $\omega$ between the ascending node to the periapsis \cite{goldstein,poisson2014gravity}. The angle $\alpha$ describes the azimuthal direction of $\hat{L}$ around $\hat{S}$ --- this corresponds in terms of standard orbit elements to the longitude of ascending node $\Omega$. The angle $\lambda$ is the inclination of the orbit, i.e. the angle between $\hat{S}$ and $\hat{L}$. Finally, $\Phi$ is the mean anomaly with respect to the pericenter passage.

As mentioned, we neglect the structure  of the smaller object (i.e. its spin and other multipoles), but the larger object's multipoles feature importantly in our analysis. Its dimensionless spin magnitude is $\tilde S_1=S_1/M^2$, which for Kerr lies between 0 (unspinning, Schwarzschild) and 1 (extremally rotating Kerr). 
Then, $\tilde M_2$ is defined here as the amount that the dimensionless quadrupole moment $M_2/M^3$ deviates from the Kerr value $(M_2)_{\text{Kerr}}/M^3 = -(S_1)^2/M^4$.\footnote{Note that this is slightly different from the parametrization of Barack \& Cutler \cite{Barack:2006pq}, who took the (entire) dimensionless quadrupole moment $M_2/M^3$ as parameter instead.} Finally, $\tilde S_2=S_2/M^3$ and $\tilde M_3=M_3/M^4$ are the dimensionless lowest-order odd parity multipoles that break equatorial symmetry, which of course vanish for Kerr.

\subsection{Evolution equations}

The five extrinsic parameters $(\theta_S,\phi_S,\theta_K,\phi_K,D)$ define the distance and orientation between the source and the solar system and are constant during the inspiral. We further take the six parameters that give the masses $\mu,M$ and the multipoles $\tilde S_i,\tilde M_i$ to be constants as well --- a good approximation if the central object is much larger than the smaller one. We will also assume everywhere that $\mu/M\ll 1$, and work to leading order in the ratio $\mu/M$.

We are left with the six parameters $\Phi, \nu,\tilde\gamma,\alpha, e,\lambda$, which describe the orbit of the smaller object in coordinates relative to the large BH spin $\hat{S}$. Following Barack \& Cutler \cite{Barack:2003fp,Barack:2006pq}, we approximate $\lambda$ to be constant --- this is known to be a good approximation \cite{Hughes:1999bq} (see also especially footnote 2 in \cite{Barack:2003fp} for further justification). We have also explicitly checked that adding in time-evolution of $\lambda$ does not change our main results; see Section \ref{sec:checks}.

We must then specify how the five orbital elements $\Phi, \nu,\tilde\gamma,\alpha, e$ evolve in time. We are interested in time-scales comparable to the radiation time-scale, which is much larger than the time-scale of individual orbits. This means we can consider evolution equations which take into account (only) the averaged, secular change of these orbit elements.

The secular change of these (Newtonian) parameters can roughly be divided in two parts: the Newtonian corrections which arise due to the non-zero multipoles (here, we consider $S_1,M_2,S_2,M_3$), and (mixed-order) post-Newtonian corrections. The former are conservative in nature and include standard effects such as Lense-Thirring precession due to $S_1$. The latter include dissipative effects on the orbital frequency $\nu$ and eccentricity $e$ due to quadrupolar radiation,\footnote{Although note that we only take into account the effects of $S_1$ on the orbit when considering the radiation corrections to $e$.} and the general relativistic precession of the angle of the periapsis $\tilde\gamma$. Note that the dissipative effects are also dependent on the central object's multipole moments since these affect the (Newtonian) orbit that is used to calculate the average quadrupolar radiation. See appendix \ref{app:oscel} (and \cite{Barack:2003fp}) for more details. The evolution equations we use are then:
\begingroup
\allowdisplaybreaks
\begin{align}
 \label{eq:evPhi} \frac{d\Phi}{dt} &= 2\pi\nu, \qquad v := (2\pi M \nu)^{1/3}\, , \qquad \cE:= (1-e^2)^{-1/2},\\
\label{eq:evnu}  \frac{d\nu}{dt} &= \frac{96}{10\pi}\frac{\mu}{M^3} v^{11}\cE^9\left\{ f_{\nu,0}(e)   +v^2 f_{\nu,1}(e)  - v^3  \cE \tilde S_1 f_{\nu,S_1}(e) \cos\lambda \right.\\
\nonumber & \left. - v^4 \cE^4  \tilde M_2 \left[f_{\nu,M_2,1}(e)(1+3\cos{2\lambda}) +  e^2 f_{\nu,M_2,2}(e)\sin^2{\lambda} \cos{2 \tilde{\gamma}}\right] \right.\\
\nonumber & \left. -v^5 \cE^3 \tilde S_2  e^{-1}  f_{\nu,S_2}(e)\sin{2\lambda} \sin{\tilde\gamma} \right.\\
\nonumber & \left. -v^6 \cE^6\tilde M_3 e^{-1}  \sin{\lambda} \left[  f_{\nu,M_3,1}(e) (3+5\cos{2\lambda})\sin{\tilde\gamma}  +e^4 f_{\nu,M_3,2}(e) \sin^2{\lambda} \sin{3 \tilde\gamma}  \right]\right\},\\
\label{eq:evgamma}\frac{d\tilde\gamma}{dt} &= 6\pi\nu v^2\cE^2\left[ 1+\frac14v^2\cE^2(26-15e^2)\right]\\
\nonumber &+ \pi\nu\left\{ -12 v^3 \cE^3\tilde S_1  \cos\lambda -\frac34 v^4 \cE^4\tilde M_2 (3+5\cos2\lambda) \right.\\
\nonumber &\left. +6  v^5\cE^5 \tilde S_2 e^{-1}  \left[(1+5e^2) \cos{2\lambda} - (1+3e^2)\right]\cot{\lambda} \sin{\tilde\gamma}\right.\\
\nonumber &\left.  +\frac{3}{32}  v^6 \cE^6 \tilde M_3e^{-1}   \left[(5+35e^2) \cos{4\lambda} - 4 \cos{2 \lambda} - (1+3e^2)\right]\csc{\lambda} \sin{\tilde\gamma} \right\},\\
\label{eq:evalpha} \frac{d\alpha}{dt} &= \pi\nu\left\{ 4 v^3 \cE^3\tilde S_1  + 3 v^4 \cE^4\tilde M_2\cos\lambda \right.\\
\nonumber &\left. -12 v^5 \cE^5\tilde S_2 e  \csc{\lambda} \cos{2 \lambda} \sin{\tilde\gamma} -\frac{3}{8}v^6 \cE^6\tilde M_3 e \cot{\lambda}  (-7+15 \cos{2 \lambda}  ) \sin{\tilde\gamma} \right\},\\
\label{eq:eve} \frac{de}{dt} &= - \frac{e}{15}\frac{\mu}{M^2}v^8\cE^7 \left\{ (304+121e^2)(1-e^2)(1+12v^2) \right. \\
\nonumber & \left. -\frac{1}{56} v^2( (8)(16705)+(12)(9082)e^2-25211e^4) \right.\\
\nonumber & \left. -  v^{3}\cE\tilde S_1\left( 654 + 6000e^2 +  \frac{789}{2}e^4    \right)\cos\lambda\right\}\\
\nonumber & + \pi\nu\left\{ 6  v^5\cE^3 \tilde S_2  \sin{2 \lambda} \cos{\tilde\gamma}  +\frac{3}{8} v^6\cE^4 \tilde M_3  \sin{ \lambda}(3+5\cos{ 2\lambda}) \cos{\tilde\gamma} \right\}.  
\end{align}
\endgroup
The functions $f_i(e)$ are simple polynomials in the (squared) eccentricity $e^2$, which we give in appendix \ref{app:oscel}.

We derived the Newtonian effects of the multipoles $S_i,M_i$ using the method of osculating elements; for more details, see appendix \ref{app:oscel}. Note that our $S_1,M_2$-dependent terms are different from those used by Barack \& Cutler \cite{Barack:2003fp,Barack:2006pq} (although our $S_1$ terms agree with Ryan \cite{Ryan:1995xi}, which was not true for \cite{Barack:2003fp,Barack:2006pq}); this is also explained in appendix \ref{app:oscel}. We explicitly checked that our differing evolution equations do not change any of the conclusions or main quantitative estimates of \cite{Barack:2003fp,Barack:2006pq}. Note further that the $\tilde S_2,\tilde M_3$ terms in $de/dt$ are non-dissipative, but arise simply from the non-conservation (on a Newtonian level) of the orbital angular momentum when the multipoles $S_2,M_3$ are non-vanishing.\footnote{The multipoles $S_1,M_2$ also break conservation of orbital angular momentum, but actually conserve this angular momentum when averaged over orbits.} In principle, $de/dt$ should also have dissipative terms proportional to $M_2,S_2,M_3$ which we neglect (as in \cite{Barack:2006pq} for $M_2$). Finally, we note that the dissipative terms in $d\nu/dt$ and $de/dt$ are correct up to 3.5PN order (i.e. one order higher than 2.5PN, which is the leading order radiation reaction); see \cite{Barack:2003fp}. At the order that the $\tilde S_2,\tilde M_3$ dissipative terms in $d\nu/dt$ arise, however, there should then also in principle be additional competing terms at the same order. These come from, for instance, higher-order multipolar radiation such as octupolar radiation. We describe a test in Section \ref{sec:checks}, inspired by \cite{Barack:2006pq}, which nevertheless indicates the robustness of our analysis even though we are not strictly operating at a consistent PN order (as is characteristic for kludge models).

Finally, we will set initial conditions for the system of evolution equations at the approximate Schwarzschild last stable orbit

\begin{equation}
	\nu_0 = (2 \pi M)^{-1} \left(\frac{1-e_0^2}{6+2e_0}\right)^{3/2}
\end{equation}

This is conservative in that it is generally expected to lead to an underestimate of the parameter estimation accuracies \cite{Gair:2017ynp}.

\subsection{Waveform and signal analysis}

With a model for the orbital evolution in hand, we construct the observed waveform based on the quadrupole approximation for Newtonian binaries following Peters and Matthews \cite{peters1963gravitational, peters1964gravitational}. Concretely, this means the metric perturbation is given by

\begin{equation}
	h_{ij} = \frac{2}{D}(P_{ik}P_{jl}-\frac{1}{2} P_{ij}P_{kl})\ddot{I}^{k l}
\end{equation}

where $P_{ij} = \eta_{ij}-n_i n_j$ with $n$ the direction to the source, and the inertial tensor, $I^{ij} = \mu r^i r^j$, is expanded in harmonics of $\nu$ as in \cite{peters1963gravitational} --- see also \cite{Barack:2003fp}.

To extract the LISA response from this waveform, we project it onto the time-dependent LISA antenna pattern functions, which are found for instance in \cite{Cutler:1997ta}. Importantly, this means we now do take into account the motion of LISA, in contrast to Sec. \ref{sec:circular}. These choices, as well as an implementation of Doppler modulation due to the motion of LISA and a mode by mode reweighing for a time-domain computation of the Fisher matrix \eqref{eqn:fisher}, are thus kept the same as in the work of Barack \& Cutler \cite{Barack:2003fp,Barack:2006pq}. We have based our implementation of these on code shared by the Black Hole Perturbation Toolkit \cite{BHPToolkit}, which in turn was modified from this original work. The main difference in our signal analysis is that we have used an updated noise curve \cite{Robson:2018ifk} (as also used in Sec. \ref{sec:circular}). 


\subsection{Numerical and calculational checks}\label{sec:checks}
We mention here a number of robustness checks that we performed on our calculations. For all of these checks, we take as baseline simulation parameters those given in Table \ref{tab:maintab}, unless otherwise specified.

In our main analysis, we keep the inclination angle $\lambda$ fixed, as in \cite{Barack:2003fp,Barack:2006pq}. As mentioned above, dissipative effects that alter $\lambda$ during the evolution are small and can be ignored \cite{Hughes:1999bq,Barack:2003fp}. However, for non-zero $\tilde S_2,\tilde M_3$, there are also Newtonian precession effects that alter $\lambda$ at a lower order than the dissipative effects. The corresponding evolution equation (\ref{eq:evlambda}) is given in the appendix. We checked that including $\lambda$ as a dynamical variable, using (\ref{eq:evlambda}) to evolve it along the trajectory, does not alter our results significantly. Specifically, for $\lambda=\{\pi/6,\pi/3,2\pi/3,5\pi/6\}$ and the other parameters chosen as in Table \ref{tab:maintab}, the result $\Delta\tilde M_1$ changes at most by $\sim 2\times 10^{-5}$.

The presence of the $1/e$ terms in the evolution equations may seem worrying for low-eccentricity orbits.\footnote{The $\sim \lambda^{-1}$ divergences in the evolution equations could similarly be worrying. However, we keep $\lambda$ fixed for most of our analyses, and moreover checked that varying $\lambda$ according to its Newtonian evolution equation (\ref{eq:evlambda}) does not qualitatively affect the results.} Of course, these terms are physical, and the $e\rightarrow 0$ divergence is an artifact of the ill-suitedness of the orbital elements to describe circular orbits. As long as $\tilde S_2/e$ and $\tilde M_3/e$ remain small, we expect our linear-order multipole analysis to remain valid.
Nevertheless, we have also checked that our evolution equations and their results are robust despite the $1/e$ terms, by redoing the analysis with the evolution equations modified by deleting (by hand) (a) the $1/e$ terms in $d\nu/dt$ (which are dissipative), (b) the $1/e$ terms in $d\nu/dt$ \emph{and} the Newtonian precession $1/e$ terms in $d\tilde\gamma/dt$. We performed both checks (a) and (b) for the simulation parameters in Table \ref{tab:maintab} and for eccentricities $e=0.1, 0.01, 0.3$. The check (a) gives barely any change at all; by contrast, the error deteriorates to $\Delta \tilde M_1\sim \mathcal{O}(10^{-1})$ for (b) --- this shows that the Newtonian precession effect on $\tilde\gamma$ is relatively important to distinguish equatorial symmetry breaking.

We also repeated the ``post-Newtonian robustness check'' of \cite{Barack:2006pq}, checking that deleting the highest PN-order dissipative terms in the evolution equations (i.e. the $\mathcal{O}(v^{13})$ term in (\ref{eq:evnu}), the $\mathcal{O}(v^4\tilde M_2^0)$ terms in (\ref{eq:evgamma}), and the $\mathcal{O}(v^{10})$ term in (\ref{eq:eve})) does not affect our results significantly. For example, repeating the simulation of Table \ref{tab:maintab} without these higher-order PN terms, the measurement accuracy on the multipoles $\Delta \tilde S_1,\Delta \tilde M_2,\Delta \tilde M_1$ changes at most by a factor of two.

Finally, we have implemented the analysis with higher precision arithmetic in order to be able to confidently establish the convergence of the numerical derivatives used to compute the Fisher information matrix \eqref{eqn:fisher}. In addition, it is well-known that such  Fisher matrices are typically not well-conditioned. It is not uncommon to find condition numbers of order $10^{19}$, although this can generally be reduced by judicious rescalings. Therefore, we have checked that the inversion is robust with respect to, for instance, the variations induced when varying the stepsize of the numerical derivative. The same is true when using a different inversion method, based on singular-value decomposition. An exception is when parameters are strongly degenerate (e.g. the angles $\theta_K,\phi_K,\lambda$ when $S \to 0$), but even then, the inversion is robust for other parameters such as the masses and the multipoles.

\section{Measuring Equatorial Symmetry Breaking with LISA}\label{sec:measure}

The ``analytic kludge'' formalism for EMRIs, pioneered by Barack and Cutler \cite{Barack:2003fp,Barack:2006pq}, and expanded here by us to include effects of odd parity multipoles $S_2,M_3$ that break equatorial symmetry, was introduced in the previous Section. We now use this formalism to simulate candidate EMRI events of which LISA detects 1 year of data before the final plunge. These simulations and their Fisher analysis then give us the accuracy to which we estimate LISA will be able to detect or constrain equatorial symmetry breaking of the central, supermassive black hole.

We introduce a single dimensionless parameter $\tilde M_1$ to parametrize equatorial symmetry breaking:
\be \tilde M_1:=\tilde S_2=\tilde M_3. \ee
In this way, we artificially ``link'' the value of the two odd parity multipoles $S_2,M_3$ to a single value --- this is actually relatively natural if the breaking of equatorial symmetry is due to the black hole having effectively a ``gravitational dipole moment'' such as in string theory black holes \cite{Bena:2020see,Bena:2020uup,Bah:2021jno}. A prototypical result for the measurement accuracies of some of the EMRI parameters is given in Table \ref{tab:maintab} (see the caption for the specific simulation parameters used). We always take the reference simulation parameters, at which \eqref{eqn:fisher} is computed, of the large black hole to be those of Kerr, i.e. $\tilde M_2 = M_2/M^3+\tilde S_1^2 = 0$ and $\tilde S_2=\tilde M_3=0$.

\renewcommand{\arraystretch}{1.3}
\begin{table}[ht]\centering
\begin{tabular}{c|c|c|c}
	\hline\hline
$\Delta (\ln \mu)$ & $\Delta (e_0)$ & $\Delta (\cos\lambda)$ & $\Delta \tilde\gamma_0$ \\
\hline
$7.4\times 10^{-4}$ & $3.9\times 10^{-4}$ & $9.04\times 10^{-3}$ & $2.4\times 10^{-1}$  \\
\hline\hline
$\Delta (\ln M)$ & $\Delta \tilde S_1 $ & $\Delta \tilde M_2 $ &  $\mathbf{\Delta \tilde M_1}$\\
\hline
$8.2\times 10^{-4}$ & $7.0\times 10^{-4}$ & $4.4\times 10^{-3}$ & $\mathbf{1.8\times 10^{-2}}$\\
\hline
\end{tabular}
\caption{The measurement accuracies for a number of parameters related to the smaller object's orbit ($\mu, e_0,\cos\lambda,\gamma_0$) and the properties of the larger black hole ($M,\tilde S_1,\tilde M_2,\tilde M_1$) for an SNR of $30$. The large black hole parameters for this simulation are $M=10^6 M_\odot, \tilde S_1 = S_1/M^2 = 0.25, \tilde M_2 = M_2/M^3+\tilde S_1^2 = 0$ and the equatorial symmetry breaking parameter is set to $\tilde M_1 = 0$. The small black hole's orbit parameters are given by $\mu = 1 M_\odot, e_0 = 0.1$, together with the angles $\tilde\gamma_0=\alpha_0=\Phi_0=0$ and $\lambda=\pi/3$. The other angles are $(\theta_S,\phi_S,\theta_K,\phi_K) = (2\pi/3, 5\pi/3, \pi/2,0)$.}
\label{tab:maintab}
\end{table}

For the particular simulation in Table \ref{tab:maintab} at a signal-to-noise ratio (SNR) of 30, we project \textbf{LISA can measure or rule out equatorial symmetry breaking to $\sim 1.8\%$}. Note that an SNR of $\sim 30$ corresponds roughly to the estimated detection treshold for EMRIs \cite{Gair:2004iv}, although in ideal conditions $\text{SNR}\sim 15$ might be sufficient \cite{babak2010mock}. Therefore, our estimate covers the weaker end of detectable signals and could actually be improved by an order of magnitude for a lucky ``golden'' EMRI.  In addition, our result is rather robust; the rest of this Section is dedicated to discussing how (little) it changes when the simulation parameters are varied.\footnote{We will not discuss varying the unimportant initial parameters $\tilde\gamma_0,\Phi_0,\alpha_0$. The distance is always fixed by setting $\text{SNR}=30$.}

The choice of angles $\theta_S,\phi_S,\theta_K,\phi_K,\lambda$ do not qualitatively alter the result for $\Delta \tilde M_1$. We confirmed this by keeping all parameters fixed to their values as in Table \ref{tab:maintab} and varying these angles one by one, considering (inspired by the values analyzed in \cite{Barack:2006pq}): $\theta_S=\{\pi/6, \pi/2\}$, $\phi_S= \{0,\pi/3\}$, $\theta_K=\{\pi/20,3\pi/4\}$, $\phi_K=\{\pi/2,\pi\}$, and $\lambda=\{\pi/6,2\pi/3,5\pi/6\}$. Varying these angles affected the result $\Delta \tilde M_1$ (compared to that of Table \ref{tab:maintab}) at most by a factor of two.

\begin{figure}[ht]\centering
 \begin{subfigure}{0.48\textwidth}
  \includegraphics[width=\textwidth]{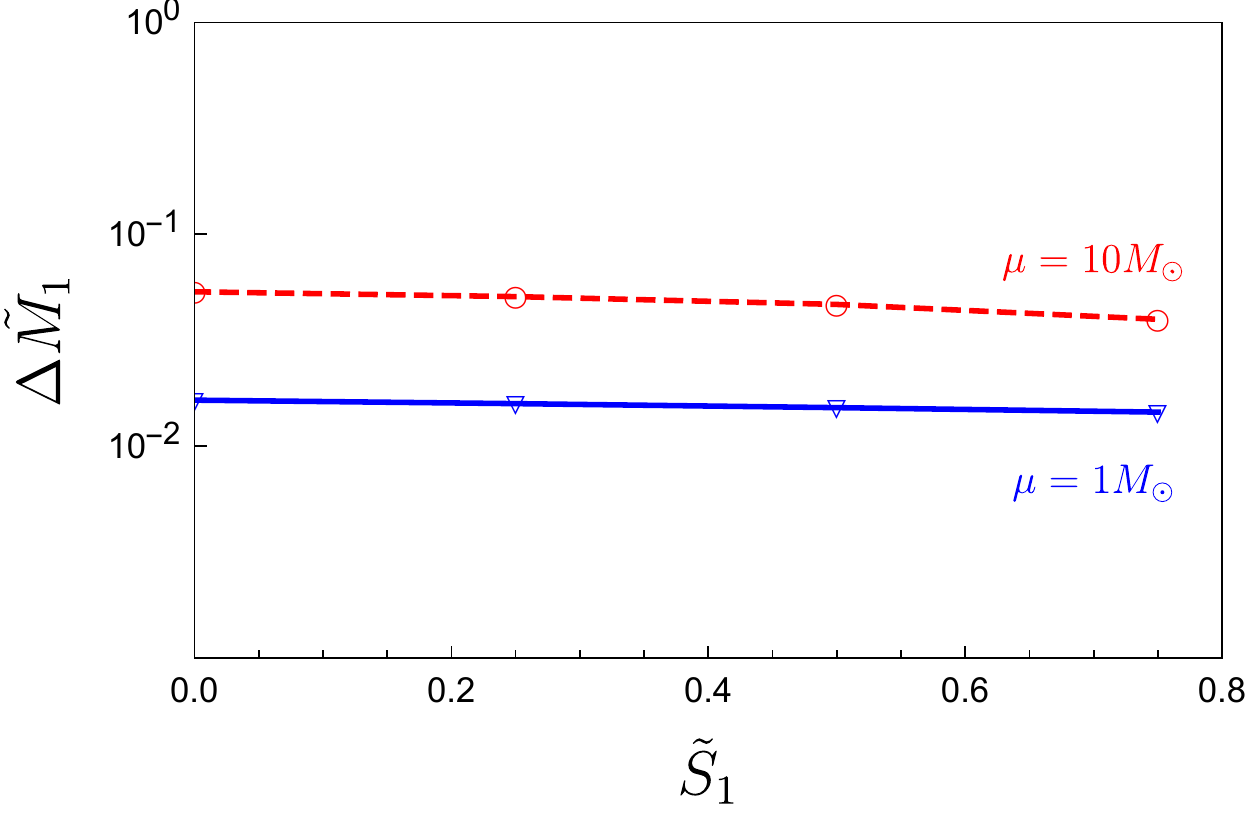}
  \caption{$M=10^5 M_\odot$}
 \end{subfigure}
 \begin{subfigure}{0.48\textwidth}
  \includegraphics[width=\textwidth]{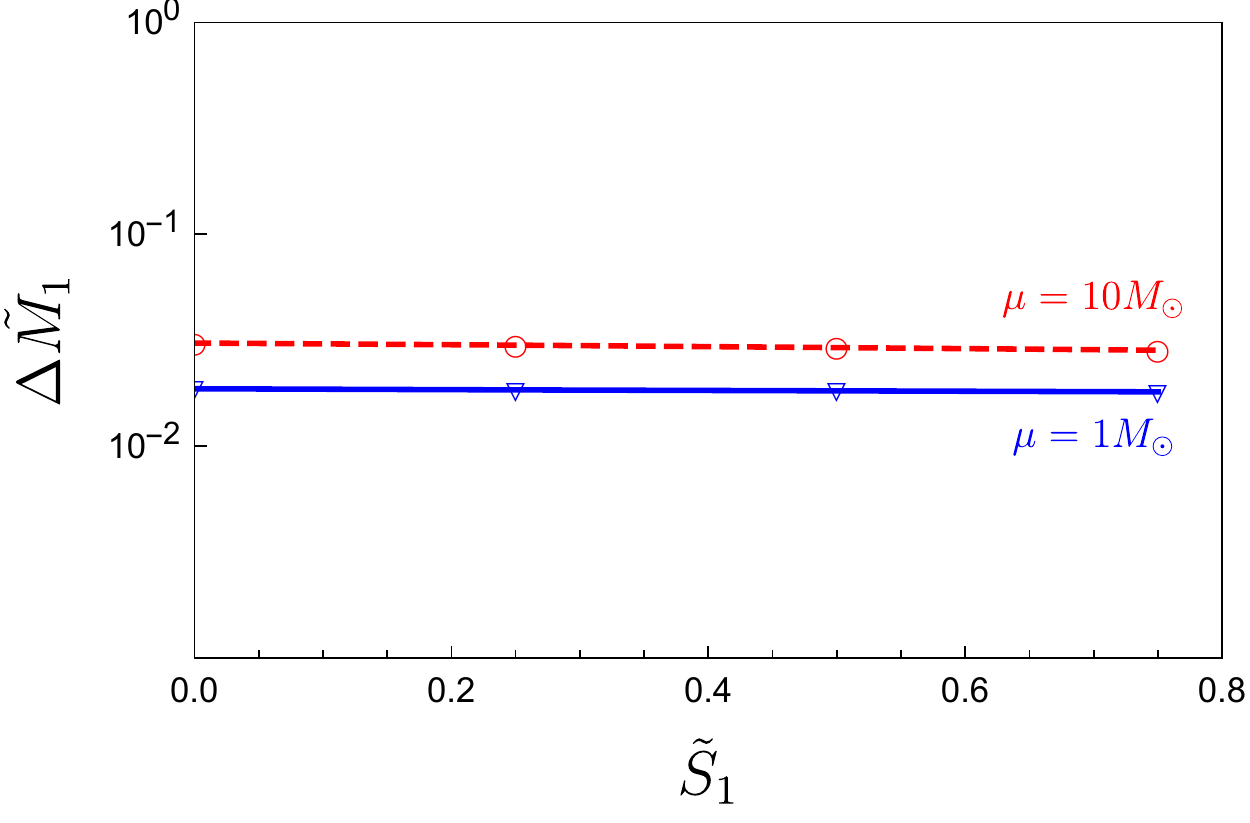}
  \caption{$M=10^6 M_\odot$}
 \end{subfigure}
 \caption{The measurement accuracy $\Delta \tilde M_1$ for equatorial symmetry breaking, for varying values of $M,\mu, \tilde S_1$; with $e_0=0.1$ (and all other parameters kept fixed to their values of Table \ref{tab:maintab}).} 
 \label{fig:varS}
\end{figure}

\begin{figure}[ht]
 \centering
 \includegraphics[width=0.48\textwidth]{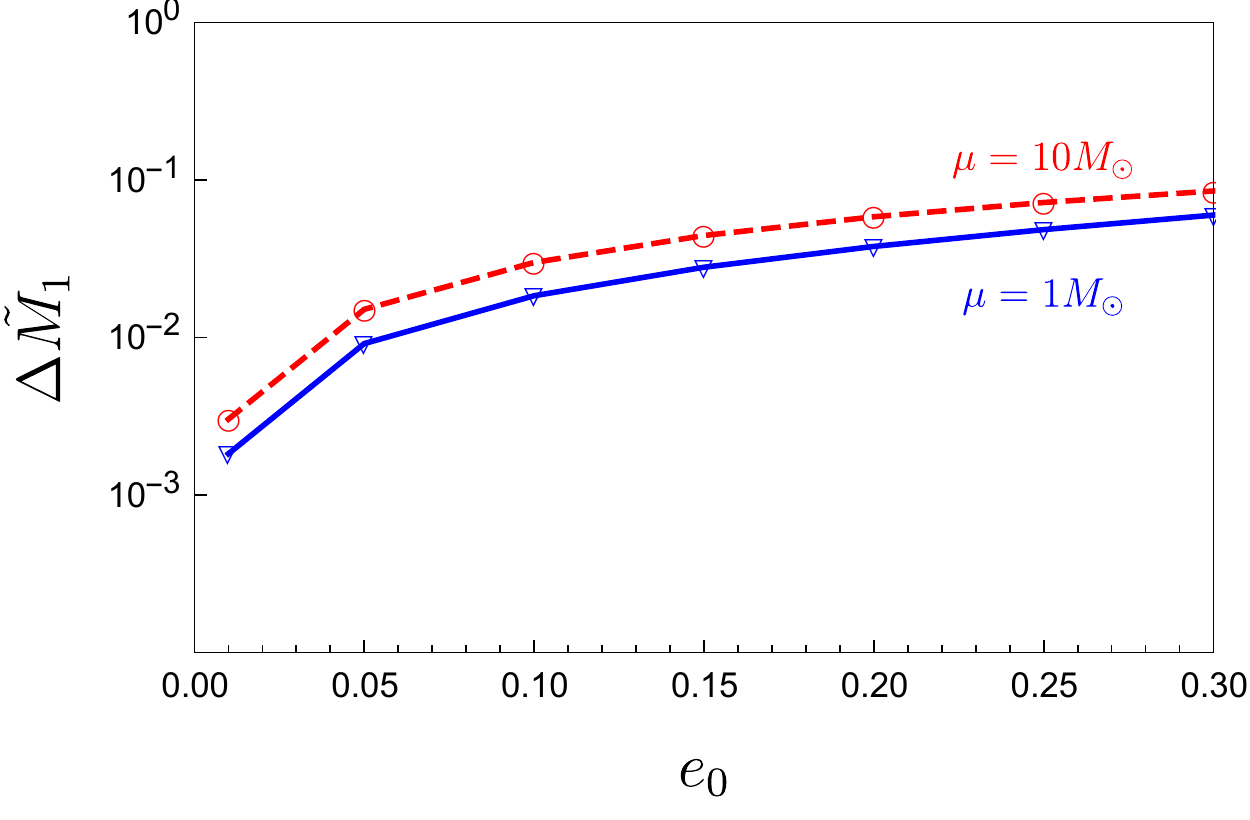}
 \caption{The measurement accuracy $\Delta \tilde M_1$ for equatorial symmetry breaking, for varying values of $\mu, e_0$; with $M=10^6 M_{\odot}$ and $\tilde S=0.25$ (and all other parameters kept fixed to their values of Table \ref{tab:maintab}).}
 \label{fig:vare}
\end{figure}

 We considered the effects of varying the parameters $M,\mu,e_0,\tilde S$; we simulated all combinations of the values $M=\{10^5,10^6\}M_\odot$, $\mu=\{1,10\}M_\odot$, $e_0=\{0.01,0.1,0.3\}$, $\tilde S_1=\{0,0.25,0.5,0.75\}$, except certain $e_0=0.3$ trajectories which led to too high eccentricities along the trajectory. How the measurement accuracy varies with these parameters is summarized in Figures \ref{fig:varS} and \ref{fig:vare}. As we see from Fig. \ref{fig:varS}, the accuracy $\Delta\tilde M_1$ is essentially insensitive to varying $\tilde S_1$, and $\mu$ and $M$ affect the result minimally.  
 
 The (initial) eccentricity $e_0$ seems to have the largest effect on $\Delta\tilde M_1$, as shown in Fig. \ref{fig:vare} (where we have also included additional data points): low initial eccentricities $\sim 0.01$ lead to much better measurement accuracies $\Delta \tilde M_1\sim 10^{-3}$.\footnote{This may seem suspect, and in particular a consequence of the $\sim 1/e$ terms in the evolution equations (\ref{eq:evPhi})-(\ref{eq:eve}). However, as mentioned in Section \ref{sec:checks}, we explicitly checked that ``leaving out'' these $1/e$ terms does not alter the result significantly. Note that, for example, the $\mathcal{O}(e^{-1})$ and the $\mathcal{O}(e)$ terms in (\ref{eq:evnu}) --- see $f_{\nu,S_2}(e)$ in (\ref{eq:fnuS2}) --- are of the same order for $e=0.1$; $3/e\approx 30$ and $(209/2) e \approx 10.5$.}
On the other hand, higher eccentricity orbits $e_0\sim 0.3$ lead to poorer measurement accuracies $\Delta \tilde M_1\sim 10^{-1}$. Note that this analytic kludge analysis is not reliable for high eccentricities \cite{Barack:2003fp,Barack:2006pq}, so we cannot extrapolate our results to arbitrary high eccentricities.

Finally, we also considered the case where $\tilde S_2,\tilde M_3$ are separate parameters, and not ``linked'' by $\tilde M_1=\tilde S_2=\tilde M_3$. We find that including $S_2$ alone yields results comparable (roughly within $10\%$) to the combined analysis. However, including independently $M_3$ can significantly worsen individual measurement accuracies. As such, in future analysis, it might be advisable to focus on $S_2$ but, rather than taking the results at face value for this particular multipole, consider it a proxy for the capability of LISA to detect equatorial symmetry breaking.

\section{Discussion}\label{sec:discussion}
From our analysis, it is clear that LISA will be able to measure and constrain equatorial symmetry breaking remarkably well in EMRIs. The effects of odd-parity (dimensionless) multipoles such as $\tilde S_2$ can be measured to within $10^{-2}$ (or better) for a large range of EMRI parameters. The conclusion is clear: LISA will give impressive measurements and constraints on equatorial symmetry breaking at the percent level!

It is very exciting that LISA will be able to measure this smoking-gun signal of beyond-GR physics. Odd-parity multipoles such as $S_2$ are higher-order, and so the naive expectation would be that measuring them very precisely would not be possible. However, we find here --- due to their special nature of breaking equatorial symmetry --- that robust and precise measurements of them are still possible. 

Given the rough nature of both the waveform used in this work (analytic kludge) as well as the data analysis technique (Fisher analysis), there is still a lot of room for potential improvement of the estimates of measurement accuracies we calculated here. First, a full Bayesian analysis should be conducted to assess the detectability of equatorial symmetry breaking, possibly with the incorporation of a realistic EMRI population model. Second, on the level of waveforms, significant improvements will likely require further assumptions about the nature of the equatorial symmetry breaking. A more complete picture of the relevant supermassive black hole spacetime is needed, given the reliance on gravitational waves from the highly relativistic region close to the horizon. Fortunately, as we have mentioned, there is no shortage of well-motivated beyond-GR models in which equatorial symmetry breaking occurs. Below, we briefly discuss how our work gives a rough estimate on how such models can be constrained. However, as mentioned, an in-depth, model-specific analysis (and one that is more suited to the highly relativistic region close to the horizon) would likely be able to improve such constraints considerably.

\paragraph{Constraining beyond-GR physics}
We mentioned in Section \ref{sec:intro} that many models of beyond-GR physics give rise to equatorial symmetry breaking. Since LISA can constrain $\tilde S_2=S_2/M^3$ to within $10^{-2}$, we can also wonder how well this would constrain such models.

For example, we can consider the family of almost-BPS black holes of \cite{Bah:2021jno}.\footnote{Parameters of other, more general string theory black holes that break equatorial symmetry \cite{Bena:2020uup} can be similarly constrained.} These have multipoles determined by a single parameter $h$; the lowest-order non-trivial multipoles $M_2, S_2$ are given by \cite{Bah:2021jno}:
\be M_2^\text{(aBPS)} = -M \left(\frac{S_1}{M}\right)^2  \left( \frac{1-h^2}{h^2}\right) ,\qquad S_2^{\text{(aBPS)}} = \mp 2 S_1 \left(\frac{S_1}{M}\right) \left( \frac{1-h^2}{h^2}\right)^{1/2} .\ee
We can set $h^{-1}=\sqrt{2}$ to fix $M_2$ to be the Kerr value $M_2=-S_1^2/M$. (Since we expect to be able to measure $M_2/M^3$ to within $10^{-2}-10^{-4}$ \cite{Barack:2006pq}, this is a reasonable first analysis.) In this case, we find:
\be \frac{S_2^{\text{(aBPS)}}}{M^3} = \mp 2 \frac{S_1^2}{M^4}.\ee
Our results then imply that such black holes can be ruled out (or detected) with LISA EMRIs as long as the central object has a spin $\tilde S_1=S_1/M^2\gtrsim 0.1$!

Other beyond-GR models can be similarly constrained. For example, higher-derivative corrections to GR can be divided into even-parity and odd-parity corrections. As mentioned in Section \ref{sec:intro}, the odd-parity corrections give rise to equatorial symmetry-breaking corrections to the Kerr metric. In terms of multipoles, to linear order in the higher-derivative correction parameters, only the even-parity corrections contribute to deviations of $M_2$ from its Kerr value, while only the odd-parity corrections contribute to $S_2\neq 0$. The analytic kludge EMRI estimate of measuring $M_2$ as discussed by Barack and Cutler \cite{Barack:2006pq}, and measuring $S_2$ as discussed here, then give a rough estimate of constraining both types of higher-derivative corrections \cite{upcomingHDmultipoles}. However, note that the analytic kludge does not capture the dynamical modifications to gravity due to the higher-derivative corrections, so a more model-specific, dynamical analysis is necessary (such as in \cite{Endlich:2017tqa}) for a more accurate estimate of their constrainability. This would also be necessary to estimate the potential measurability of the non-zero $S_4$ induced in dCS gravity \cite{Sopuerta:2009iy}.

\section*{Acknowledgments}
We would like to thank F. Bacchini, I. Bah, I. Bena, A. Cardenas-Avendano, D. Gates, P. Heidmann, T. Hertog, L. K\"uchler, T. Li, Y. Li, B. Ripperda, B. Vercnocke, N. Warner for interesting discussions. We would also like to thank F. Sevenants for his always efficient and patient support. This work makes use of the Black Hole Perturbation Toolkit. DRM is supported by ERC Advanced Grant 787320 - QBH Structure, ERC Starting Grant 679278- Emergent-BH, and FWO Research Project G.0926.17N. K.F. is Aspirant FWO-Vlaanderen (ZKD4846-ASP/18). This work is also partially supported by the KU Leuven C1 grant ZKD1118 C16/16/005. 

\appendix

\section{Details of Near-Circular, Near-Equatorial Analysis}\label{app:circular}

In this appendix, we provide further details for the near-circular, near-equatorial analysis discussed in Sec. \ref{sec:circular}. Effectively, this will follow the seminal work of Ryan \cite{Ryan:1995wh,Ryan:1995zm,Ryan:1997hg} with the crucial difference that reflection symmetry across the equatorial plan is not imposed.

First, in absence of radiation reaction, one can determine, in a large separation expansion, the properties of near-equatorial circular orbits by studying geodesics in an axisymmetric spacetime of a specified multipole structure
\begin{equation}\label{eqn:metric}
	d s^2 = -F(r,\xi)(dt - \omega(r,\xi) d \phi)^2 + \frac{1}{F(r,\xi)}\left( e^{2\gamma(r,\xi)}(dr^2 + r^2 d\xi^2) + r^2 \sin^2{\xi} d\phi^2 \right) \, .
\end{equation}

Similar to \cite{Ryan:1995wh}, we find it most convenient to determine the functions $F$, $\gamma$, $\omega$ starting from a complex function $\tilde{\Xi}$ related to the Ernst potential $\cE$ \cite{ernst1968new}
\begin{equation}
	\cE = F + i \psi = \frac{r - \tilde{\Xi}}{r + \tilde{\Xi}} \, ,
\end{equation}
where the real part of the Ernst potential is the $F$ appearing in \eqref{eqn:metric}. One then has \cite{Ryan:1995wh}
\begin{equation}
	\omega = -\int^{\infty}_{\rho} \frac{\rho'}{F^2} \frac{\partial \psi}{\partial z} d \rho' \, ,
\end{equation}
and
\begin{equation}
	\gamma = \frac{1}{4} \int^{\infty}_{\rho} \frac{\rho'}{F^2} \left((\frac{\partial F}{\partial \rho})^2 +(\frac{\partial \psi}{\partial \rho})^2 -(\frac{\partial F}{\partial z})^2 -(\frac{\partial \psi}{\partial z})^2  \right) d \rho' \, ,
\end{equation}
using $z = r \cos{\xi}$ and $\rho = r \sin{\xi}$. Now, expanding $\tilde{\Xi}$ as\footnote{At this point reflection symmetry would be characterized by reality of $a_{j(2k)}$ while $a_{j(2k+1)}$ are purely imaginary \cite{meinel1995asymptotically, kordas1995reflection}.} 
\begin{equation}
	\tilde{\Xi} = \sum^{\infty}_{j, k = 0} a_{j k} \frac{\sin^{j}{\xi}\cos^k{\xi}}{r^{j+k}} \, ,
\end{equation}
all coefficients $a_{jk}$ are determined in terms of $a_{j0}$ and $a_{j1}$ by means of the recurrence relation (24) in \cite{Ryan:1995wh}, see also \cite{fodor1989multipole, sotiriou2004corrections, fodor2021calculation}. Finally, the Ernst potential can in turn be related to the multipole moments, in the formulation of Geroch-Hansen \cite{geroch1970multipole,hansen1974multipole, fodor1989multipole}, which can thus be fixed by the coefficients $a_{jk}$. We do not repeat the procedure here; it can again be found in Sec. III (D) of \cite{Ryan:1995wh}, or is alternatively described in e.g. \cite{fodor1989multipole, fodor2021calculation}.

Having fixed \eqref{eqn:metric} in this way in terms of its multipole moments up to the desired order, one can then straightforwardly look for solutions of the geodesic equation with constant $r$ and $\xi$. Expressing those in terms of the frequency $2 \pi \nu = \frac{d \phi}{d t}$ or equivalently $v= (2 \pi \nu M)^{1/3}$, one finds the results quoted in the main text \eqref{eqn:xi} and
\begingroup
\allowdisplaybreaks
\begin{align} \label{eqn:fullcircr}
	r &= M v^{-2} \left\lbrace 1 -v^2 -\frac{2 S_1}{3 M^2}v^3 +v^4 \left(-\frac{1}{2}-\frac{M_2}{2 M^3} \right) + v^6 \left(-\frac{1}{2}-\frac{5 M_2}{6 M^3}-\frac{4 S_1^2}{9 M^4}   \right)  \right\rbrace \nn \\
	&+ M v^{-2} \left\lbrace - v^7 \left( \frac{S_1}{3 M^2} + \frac{4 M_2 S_1}{3 M^5} - \frac{S_3}{M^4}  \right) + v^8 \left( -\frac{5}{8} - \frac{25 M_2}{14 M^3} - \frac{3 M_2^2}{4 M^6} + \frac{5 M_4}{8 M^5} - \frac{29 S_1^2}{28 M^4}\right)  \right\rbrace  \nn \\
	&- M v^7 \left( \frac{2 S_1}{3 M^2} + \frac{40 M_2 S_1}{9 M^5} + \frac{40 S_1^3}{81 M^6} - \frac{10 S_3}{3 M^4}\right) \nn \\
	&+ M v^8 \left( -\frac{7}{8} - \frac{25 M_2}{6 M^3} - \frac{19 M^2_2}{6 M^6} + \frac{21 M_4}{8 M^5} - \frac{115 S_1^2}{36M^4} - \frac{28 M_2 S_1^2}{9 M^7} + \frac{19 S_2^2}{2 M^6} + \frac{8 S_1 S_3}{3 M^6}\right) \nn \\
	&- M v^9\left(\frac{11 S_1^3}{6 M^6}+\frac{9
		M_2^2 S_1}{2 M^8}+\frac{73 M_2 S_1}{6 M^5}-\frac{5
		M_4 S_1}{2 M^7}+\frac{5 S_1}{4 M^2}-\frac{12 M_3
				S_2}{M^7}-\frac{3 M_2 S_3}{M^7}-\frac{53 S_3}{6
		M^4}+\frac{5 S_5}{4 M^6}\right) \nn \\ 
	&+M v^{10} \left(-\frac{160 S_1^4}{243
		M^8}-\frac{257309 M_2 S_1^2}{16632 M^7}-\frac{471 S_1^2}{56
		M^4}+\frac{2609 S_3 S_1}{198 M^6}-\frac{11 M_2^3}{6
		M^9}-\frac{6980 M_2^2}{693 M^6}\right) \nn \\ 
	&+M v^{10} \left( \frac{15 M_3^2}{4 M^8}+\frac{23789
				S_2^2}{792 M^6}-\frac{9899 M_2}{1008 M^3}+\frac{5 M_2
		M_4}{2 M^8}+\frac{4355 M_4}{528 M^5}-\frac{35 M_6}{48
		M^7}-\frac{21}{16}\right) \, .
\end{align}
\endgroup

Similarly, one can find the associated energy and angular momentum (the conserved charges associated to the Killing vectors $\partial_t$ and $\partial_{\phi}$). For easy comparison with \cite{Ryan:1995wh}, we instead quote the variation with frequency $\Delta E = -\nu \frac{d E}{d \nu}$
\begingroup
\allowdisplaybreaks
\begin{align} \label{eqn:fullDeltaE}
	\frac{\Delta E}{\mu} &= \frac{v^2}{3} - \frac{v^4}{2}+\frac{20 S_1}{9 M^2} v^5 + v^6\left(-\frac{27}{8}+\frac{M_2}{M^3}\right)+\frac{28 S_1}{3 M^2}v^7 + v^8 \left( -\frac{225}{16} + \frac{80 S_1^2}{27 M^4}  + \frac{70 M_2}{9 M^3}\right) \nn \\
	&+v^9 \left( \frac{81 S_1}{2 M^2} + \frac{6 M_2S_1}{M^5} -\frac{6 S_3}{M^4}\right)+ v^{10} \left( -\frac{6615}{128} + \frac{935 M_2}{24 M^3} + \frac{35 M_2^2}{12 M^6} - \frac{35 M_4}{12 M^5}+\frac{115 S_1^2}{18 M^4}\right) \nn \\
	&+ v^{11} \left( \frac{165 S_1}{M^2} + \frac{968 M_2 S_1}{27M^5}+\frac{1408 S_1^3}{243 M^6}-\frac{352 S_3}{9 M^4}\right) \nn\\
	&+ v^{12} \left( -\frac{45927}{256} + \frac{9147 M_2}{56 M^3} +\frac{93 M_2^2}{4 M^6} -\frac{99 M_4}{4 M^5} - \frac{123 S_1^2}{14 M^4} + \frac{24 M_2 S_1^2}{M^7} - \frac{57 S_2^2}{M^6} - \frac{24 S_1 S_3}{M^6}\right) \nn\\
	&+ v^{13} \left( \frac{20475 S_1}{32 M^2} + \frac{32435 M_2 S_1}{252 M^5} + \frac{260 M_2^2 S_1}{9 M^8}-\frac{325 M_4 S_1}{18 M^7} + \frac{15080 S_1^3}{567 M^6}  - \frac{6305 S_3}{36 M^4} + \frac{65 M_2 S_3}{3 M^7} \right) \nn \\
	&+ v^{13} \left( -\frac{65 M_3 S_2}{M^7} + \frac{65 S_5}{6 M^6} \right) + v^{14} \left(\frac{8624 S_1^4}{729
		M^8}+\frac{13766 M_2 S_1^2}{81 M^7}-\frac{77995 S_1^2}{432
		M^4}-\frac{161 S_3 S_1}{M^6} + \frac{385 M_2^3}{36
		M^9}\right) \nn \\ &+ v^{14} \left( \frac{100411 M_2^2}{864 M^6}+\frac{2160829 M_2}{3456 M^3}-\frac{385 M_2
		M_4}{24 M^8}-\frac{38045 M_4}{288 M^5}+\frac{385 M_6}{72
		M^7}-\frac{617463}{1024}\right) \nn \\
	&+ v^{14} \left( -\frac{77 M_3^2}{4 M^8}-\frac{18577
				S_2^2}{54 M^6} \right) \, .
\end{align}
\endgroup
(This expression can be compared to the similar, less general expression in \cite{Mukherjee:2020how} for the Kerr-NUT spacetime; note that there, $S_0\neq0$, so asymptotic flatness is broken along with equatorial symmetry.)

These results can now be further combined with the quadrupole emission \eqref{eqn:ppquadrupole}, in an adiabatic approximation, to find the frequency-domain phase $\psi(f)$ in \eqref{eqn:fdomainwave}. In particular \cite{Buonanno:2009zt}
\begin{equation}
	\psi(f) = 2 \pi f t_{*}-\phi_{*} + 2 \pi \int^{f_{*}}_f  (f -  f') \frac{d E}{d f} (\frac{d E}{d t})^{-1} df' -\frac{\pi}{4} \, ,
\end{equation}
where one should take care to use the gravitational wave frequency $f = 2\nu$, and $t_{*}$, $\phi_{*}$ are a reference time and phase. The leading order corrections due to the equatorially asymmetric multipole moments can then be found to be \eqref{eqn:oddpsi} while for the others they are given by \cite{Ryan:1997hg}
\begin{align} \label{eqn:evenpsi}
	\delta \psi_{\text{sym}} &= \frac{3}{128}(\frac{M}{\mu})(\pi M f)^{-5/3}\left(-\sum_{\text{even}  l\neq 4} \frac{(-1)^{l/2}40(2l+1)(l+1)!! (\pi M f)^{2l/3}}{3 (2l-5)(l-4)l!!} \frac{M_l}{M^{l+1}}  \right) \nn \\
	&+ \frac{3}{128}(\frac{M}{\mu})(\pi M f)^{-5/3}\left(-\sum_{\text{odd} l\geq 3} \frac{(-1)^{(l-1)/2}80(2l+5)l!! (\pi M f)^{(2l+1)/3}}{3 (l-2)(2l-7)(l-1)!!} \frac{S_l}{M^{l+1}} \right) \nn \\
	&+ \frac{3}{128}(\frac{M}{\mu})(\pi M f)^{-5/3}\left(-50 \frac{M_4}{M^5} (\pi M f)^{8/3} \ln{\pi M f}+\frac{113}{3} \frac{S_1}{M^2} (\pi M f) \right)
\end{align}
where one should take care to also include the current quadrupole radiation reaction for $S_1$.

We use the LISA sensitivity curve as given in \cite{Robson:2018ifk}
\begin{equation} \label{eqn:PSD}
	S_n(f) = \frac{10}{3 L^2} \left( 2(1+\cos^2{(f/f_*)}) \frac{S_I (f)}{(2 \pi f)^4} + S_{II}(f) \right) \left( 1+ \frac{6}{10}(\frac{f}{f_*})^2 \right) + S_c(f) \, ,
\end{equation}
with\footnote{These expressions are only used explicitly here so duplicate definitions should not cause any confusion.} $L=2.5 \text{Gm}$, $f_* = 19.09 \text{mHz}$ and
\begin{align}
	S_I (f) &= (3 \times 10^{-15} \text{m} \text{s}^{-2})^2 \left(1+(\frac{0.4 \text{mHz}}{f})^2 \right) \left(1+(\frac{f}{8 \text{mHz}})^4 \right) \text{Hz}^{-1} \, , \\
	S_{II}(f) &= (1.5 \times 10^{-11} \text{m})^2 \left(1+(\frac{2 \text{mHz}}{f})^4 \right) \text{Hz}^{-1} \, , \\
	S_c(f) &= A (f/\text{Hz})^{-7/3} e^{-(f/\text{Hz})^{\alpha} + \beta (f/\text{Hz}) \sin{(\kappa f/\text{Hz})}} \left(1+\tanh{(\gamma(f_k-f/\text{Hz}))}\right) \text{Hz}^{-1} \, ,
\end{align}
where the confusion noise $S_c(f)$ was estimated in \cite{Cornish:2017vip}. We use here the four year values of the associated parameters
\begin{equation}
	A=9\times10^{-45} \, , \quad	\alpha = 0.138 \, , \quad \beta = -221 \, , \quad \kappa=521 \, , \quad \gamma = 1680 \, ,\quad f_k=0.00113 \, .
\end{equation}

	\begin{table}[ht]
		\centering
		\begin{tabular}{c c c c c c c} 
			
			\hline
			\hline
			\rule{0pt}{3ex} $x$ & $L(\Delta t_*)$ & $L(\Delta \phi_*)$ & $L(\Delta \mu/\mu)$ & $L(\Delta M/M)$  & $L(\Delta \tilde{S}_1)$ & $L(\Delta x$)  \\ [0.5ex] 
			\hline
			
			\rule{0pt}{3ex} $\tilde{M}_2$ & $2.5$& $2.3$ & $-2.9$ & $-3.1$ & $-2.8$ & $-3.2$  \\
			
			$\tilde{S}_3$ & $2.4$& $2.3$ & $-2.9$ & $-3.1$ & $-2.8$ & $-2.8$\\
			
			$\tilde{M}_4$ & $2.5$ & $2.3$ & $-2.9$ & $-3.1$ &  $-2.8$ & $-2.2$  \\
			
			$\mathbf{\tilde{S}_2^2}$ & $3.1$ & $2.6$ & $-2.6$ & $-2.8$ &  $-2.5$ & $\mathbf{-1.0}$  \\
			
			$\tilde{S}_5$ & $2.4$ & $2.4$ & $-2.9$ & $-3.1$ &  $-2.7$ & $-1.3$ \\
			
			$\mathbf{\tilde{M}_3 \tilde{S}_2}$ & $2.4$ & $2.4$ & $-2.9$ & $-3.1$ &  $-2.7$ & $\mathbf{-2.1}$  \\
			
			$\tilde{M}_6$ & $2.5$ & $2.4$ & $-2.8$ & $-3.1$ &  $-2.7$ & $-0.4$  \\
			
			$\mathbf{\tilde{M}_3^2}$ & $2.5$ & $2.4$ & $-2.8$ & $-3.1$ &  $-2.7$ & $\mathbf{-1.0}$  \\
			\hline
			\hline
		\end{tabular}
		\caption{The errors for the different parameters in the waveform model \eqref{eqn:fdomainwave} when including multipole moments individually given 1 year of LISA observation before the ISCO for SNR=30, assuming the multipoles vanish, $M=10^5 M_{\odot}$ and $\mu = 10M_{\odot}$. We abbreviate $\log_{10}(\ldots)=L(\ldots)$ and $\tilde{S}_l = \frac{S_l}{M^{l+1}}$, $\tilde{M}_l = \frac{M_l}{M^{l+1}}$.}
		\label{tbl:ryanindividually}
	\end{table}

\section{Deriving the Analytic Kludge Evolution Equations}\label{app:oscel}
In this appendix, we give the details of how we arrived at various elements of the evolution equations (\ref{eq:evPhi})-(\ref{eq:eve}). Note that the frequency $\nu$ is essentially defined via the evolution equation (\ref{eq:evPhi}) of the mean anomaly $\Phi$, which covers an angle of precisely $2\pi$ between pericenter passages.

\subsection{Overview and discussion}
The $\leq 3.5$PN non-multipolar dissipative terms in $d\nu/dt$ (terms on first line of (\ref{eq:evnu}) proportional to $f_{\nu,0}(e)$ and $f_{\nu,1}(e)$), and $de/dt$ (first two lines of (\ref{eq:eve})), as well as the 2PN expression for $d\tilde\gamma/dt$ (first line of (\ref{eq:evgamma})) were taken directly from \cite{Barack:2003fp} and were originally given in \cite{Junker:1992kle}; we do not rederive these. The dissipative radiation terms proportional to $\tilde S_1$ (on the first line of (\ref{eq:evnu}) for $d\nu/dt$ and the third line of (\ref{eq:eve}) for $de/dt$) were derived by Ryan \cite{Ryan:1995xi}. We rederive and confirm here (see below) the dissipative $\tilde S_1$ term in (\ref{eq:evnu}) but not that in (\ref{eq:eve}), although it would be straightforward to generalize our methods to rederive this as well. Note that the $\tilde S_1$ dissipative terms we give here are indeed compatible with \cite{Ryan:1995xi}, whereas the terms in Barack \& Cutler \cite{Barack:2003fp,Barack:2006pq} are actually not: \cite{Barack:2003fp,Barack:2006pq} does not account for the shift in the definition of the parameter $a$ in \cite{Ryan:1995xi} with respect to the energy and thus the orbital frequency (see eqs. (6), (11), (14), (15) in \cite{Ryan:1995xi}).

All other terms were (re)derived by us using the method of osculating elements with multi-scale evolution; see below in Section \ref{sec:oscel:method}.
In this way, we arrive at the conservative, secular ``precession'' effects which are given in the three last lines of $d\tilde\gamma/dt$ in (\ref{eq:evgamma}), the entire expression for $d\alpha/dt$ in (\ref{eq:evalpha}), and the last line in $de/dt$ in (\ref{eq:eve}). Note that our $\tilde M_2$-contribution to $d\tilde\gamma/dt$ has a different $\lambda$ dependence than given in \cite{Barack:2003fp,Barack:2006pq}; however, our terms are consistent with \cite{Lai:1995hi}.

This method also gives Newtonian, secular effects for an average evolution $d\lambda/dt$, which we ignore in the evolution (\ref{eq:evPhi})-(\ref{eq:eve}). For completeness, we give the result here:
\be \label{eq:evlambda} \frac{d\lambda}{dt} = \pi\nu\left\{ -12 v^5 \cE^5 \tilde S_2 e^{-1}  \cos^2{\lambda} \cos{\tilde\gamma}  -\frac{3}{8} v^6\cE^6 \tilde M_3 e^{-1} \cos{\lambda} (3+5\cos{2\lambda}) \cos{\tilde\gamma} \right\}  .\ee
As discussed in Section \ref{sec:checks}, we explicitly checked that including additionally evolving $\lambda$ using (\ref{eq:evlambda}) does not alter our results appreciably.

The dissipative terms can also be calculated using the osculating elements (see below in Section \ref{sec:oscel:method}). 
In this way, we obtain the multipolar dissipative terms in (\ref{eq:evnu}) for $d\nu/dt$ (all contributions except those proportional to $f_{\nu,0}(e),f_{\nu,1}(e)$). The functions $f_i(e)$ in (\ref{eq:evnu}) are given by:
\begingroup
\allowdisplaybreaks
\begin{align}
  f_{\nu,0}(e) &= \left(1+\frac{73}{24} e^2 + \frac{37}{96} e^4\right)(1-e^2),\\
  f_{\nu,1}(e) &=  \frac{1273}{336} - \frac{2561}{224}e^2 - \frac{3885}{128}e^4 - \frac{13147}{5376}e^6,\\
  f_{\nu,S_1}(e)&=  \frac{193}{12} + \frac{647}{8}e^2 + \frac{1171}{32}e^4 + \frac{65}{64} e^6  ,\\
  f_{\nu,M_2,1}(e) &= \frac{7}{4} + \frac{821}{192}e^2 - \frac{1855}{192}e^4 - \frac{2979}{512} e^6 - \frac{59}{256}e^8,\\
  f_{\nu,M_2,2}(e) &=   \frac{897}{64} - \frac{1919}{192}  e^2 - \frac{31055}{1536}  e^4 -  \frac{79}{64}  e^6  ,\\
 \label{eq:fnuS2} f_{\nu,S_2}(e) &=  3 + \frac{209}{2}e^2 + \frac{2097}{8}e^4 + \frac{8951}{96}e^6+\frac{685}{256}e^8,\\
 \label{eq:M3finiteeguy} f_{\nu,M_3,1}(e) &=  \frac{1}{8} + \frac{373}{48}e^2 + \frac{1367}{128}e^4   - \frac{26869}{1024} e^6 - \frac{100637}{8192} e^8 - \frac{6473}{16384}e^{10} + \frac{683}{196608}e^{12}\\
 \nonumber &+ \frac{67}{49152}e^{14} +\frac{175}{262144}e^{16} + \mathcal{O}(e^{18}) ,\\
  f_{\nu,M_3,2}(e) &= \frac{18595}{768} - \frac{78145}{6144} e^2 - \frac{247775}{12288}e^4 - \frac{12455}{12288}e^6 .
\end{align}
\endgroup

The dissipative term in $d\nu/dt$ proportional to $\tilde S_1$ is not the same as that given in \cite{Barack:2003fp,Barack:2006pq}, but is precisely  equivalent to that found by Ryan \cite{Ryan:1995xi}, as mentioned above. The term in $d\nu/dt$ proportional to $\tilde M_2$ is also different than that in \cite{Barack:2006pq}: in \cite{Barack:2006pq}, this term was taken from the Kerr value given in \cite{Gair:2005ih}. However, in any case, this term from \cite{Gair:2005ih} is only accurate to $O(\lambda^0)$ (as also mentioned in \cite{Barack:2006pq}), and moreover mixes contributions from $M_2$ and $-S_1^2/M$ as these are indistinguishable in Kerr. Indeed, for near-circular, near-equatorial orbits ($e=\lambda=0$) the coefficient $33/16$ given in $d\nu/dt$ in \cite{Barack:2006pq} (eq. (5) therein) is clearly a mix of a contribution of $2$ from $M_2$ and a contribution of $1/16$ from $-S_1^2/M$ --- see e.g. eq. (55) in \cite{Ryan:1995wh}.

It is in principle possible to redefine the parameters (such as $e$) that we are using to parametrize the orbits. Such a redefinition could shift the actual coefficients appearing in the evolution equations. For example, if we shift $e\rightarrow e[1 + S_1(c_0+c_2e^2+\cdots)]$ with arbitrary constants $c_i$, then all of the $\mathcal{O}(e^2)$ and higher order coefficients would change. Such a redefinition could then be interpreted to be the source of the discrepancies between our coefficients and those of \cite{Barack:2003fp,Barack:2006pq}. However, while this reasoning could in principle apply for the $e$-dependent terms, it is reasonable to demand that the $e\rightarrow 0$ limit remains well-defined and finite. This means the $\mathcal{O}(e^0)$ term can never be altered, and also this term does not match between our expressions and those of \cite{Barack:2003fp,Barack:2006pq}. The fact that our calculations give the same expressions for the $S_1$ terms as in Ryan \cite{Ryan:1995xi}, shows that our definitions of parameters (including $e$) are compatible with those of \cite{Ryan:1995xi} (see eqs. (2) and (6) therein).

Finally, we discuss briefly some of the more peculiar features of the $S_2,M_3$ terms in the evolution equations (\ref{eq:evnu})-(\ref{eq:eve}).
We notice that the odd-parity terms in $d\nu/dt$ all have a $\sim\sin\lambda$ dependence, which is consistent with there being no such term linear in the odd parity multipoles in the near-equatorial analysis of Section \ref{sec:circular}.
The dependence on $\sim \csc\lambda$ in other evolution equations gives a divergence as $\lambda\rightarrow 0$; there are also divergences as $e\rightarrow 0$. However, these divergences are an artifact of the parametrization using orbital elements, which are not always well-suited for orbits with $\lambda\sim 0$ or $e\sim 0$. In particular, the divergence as $\lambda\rightarrow 0$ for the $S_2,M_3$ terms are simply reflecting the fact that the orbiting object must experience a finite force in the $z$ direction when its orbit is (temporarily) aligned with the equatorial plane. The divergence as $e\rightarrow 0$ can be thought of physically as if the object is ``chasing'' its own perihelion during its orbit when eccentricity is very low, resulting in a $\sim 1/e$ divergence in $d\tilde\gamma/dt$. In addition, this results in the period (which is defined between perihelion passes) diverging as $\sim 1/e$, which in turn features as a $\sim 1/e$ divergence in the radiation dissipation (in $d\nu/dt$).

\subsection{Method}\label{sec:oscel:method}
Here, we describe the method of osculating elements with multi-scale evolution. This is described in \cite{poisson2014gravity}, Sections 3.3 and 12.9; we summarize the key elements here. The following relations between the object's distance $r$ to the central object, and the parameters $p,e$ are explicitly kept fixed:
\be
	r = \frac{p}{1+e\cos{\psi}} , \qquad p = a(1-e^2),\qquad   	h = \sqrt{Mp}, \\
\ee
where $\psi$ is the true anomaly and $h$ is the orbital angular momentum. The position and velocity of the orbiting object are parametrized by $\vec{r}_{\text{Kepler}}\equiv (x,y,z)$ and $\vec{v}_{\text{Kepler}}\equiv (v_x,v_y,v_z)$ with:
\begingroup
\allowdisplaybreaks
\be \label{eq:xv}\begin{aligned}
x &= r \left(\cos \alpha \cos(\tilde\gamma + \psi) - \cos{\lambda} \sin \alpha \sin(\tilde\gamma + \psi)\right) \, , \\
y &= r \left(\sin \alpha \cos(\tilde\gamma + \psi) + \cos{\lambda} \cos \alpha \sin(\tilde\gamma + \psi)\right) \, , \\
z &= r \sin \lambda \sin(\tilde\gamma + \psi),\\
v_x &= - \sqrt{\frac{M}{p}} \left(\cos \alpha (\sin(\tilde\gamma + \psi) + e \sin \tilde\gamma) + \cos{\lambda} \sin \alpha (\cos(\tilde\gamma + \psi) + e \cos{\tilde\gamma})\right) \, , \\
v_y &= \sqrt{\frac{M}{p}} \left(\sin \alpha (\sin(\tilde\gamma + \psi) + e \sin \tilde\gamma) - \cos{\lambda} \cos \alpha (\cos(\tilde\gamma + \psi) + e \cos{\tilde\gamma})\right) \, , \\
v_z &= \sqrt{\frac{M}{p}} \sin \lambda \left(\cos(\tilde\gamma + \psi) + e \cos{\tilde\gamma}\right) \, ,
\end{aligned}\ee
\endgroup
which further define $\tilde\gamma,\alpha,\lambda$. The orbital elements $\tilde\mu^a=(p,e,\lambda, \alpha,\tilde\gamma,\psi)$ completely define the orbit. For a purely (unperturbed) Newtonian orbit, all elements except $\psi$ would be constants. Since we are perturbing the orbit (by the multipoles of the central, gravitating object), we allow these to be explicit functions of time. We then use:
\be \vec{r} = \vec{r}_{Kepler}(t,\tilde\mu^a), \qquad \vec{v} = \vec{v}_{\text{Kepler}}(t,\tilde\mu^a),\ee
to describe the orbit's position and velocity along the perturbed orbit. In order for these to be compatible, we must have:
\begin{equation}
	\sum_a \frac{\partial r_{\text{Kepler}}}{\partial \tilde\mu^a} \frac{d \tilde\mu^a}{dt} = 0, \qquad \sum_a \frac{\partial v_{\text{Kepler}}}{\partial \tilde\mu^a} \frac{d \tilde\mu^a}{dt} = f_{\text{perturb}},
\end{equation}
where $f_{\text{perturb}} = f-f_{\text{Kepler}}$ is the perturbing force. These give six constraints, entirely fixing the first order evolutions of $\mu^a$. Explicitly, the linear order equations are (see \cite{poisson2014gravity} (3.69) \& (3.70)):
\begingroup
\allowdisplaybreaks
\begin{align}\label{eq:pevol}
\frac{d p }{d \psi} &= 2 \frac{p^3}{G M} \frac{1}{(1+e \cos{\psi})^3} \mathcal{S} \, ,\\
\frac{d e }{d \psi} &= \frac{p^2}{GM} ( \mathcal{R} \frac{\sin \psi}{(1+e \cos{\psi})^2}+ \frac{2 \cos \psi + e(1+\cos^2 \psi)}{(1+e\cos \psi)^2} \mathcal{S}) \, ,\\
\frac{d \lambda }{d \psi} &= \frac{p^2}{GM} (\frac{\cos(\tilde\gamma + \psi)}{(1+e\cos{\psi})^3} \mathcal{W}) \, ,\\
\sin{\lambda} \frac{d \alpha }{d \psi} &= \frac{p^2}{GM} (\frac{\sin(\tilde\gamma + \psi)}{(1+e\cos{\psi})^3} \mathcal{W}) \, , \\
\frac{d \tilde\gamma}{d \psi}  &= \frac{1}{e} \frac{p^2}{GM} (-\mathcal{R}  \frac{\cos \psi}{(1+e\cos{\psi})^2} + \frac{2  + e \cos \psi}{(1+e\cos \psi)^3} \mathcal{S} \sin \psi - e \cot \lambda \frac{\sin(\tilde\gamma + \psi)}{(1+e\cos{\psi})^3} \mathcal{W})\label{eq:omegaevol} \, ,\\
\frac{d \psi}{dt} &= \sqrt{\frac{G M}{p^3}} (1+e\cos{\psi})^2 + \frac{1}{e} \sqrt{\frac{p}{GM}} (\mathcal{R} \cos{\psi} - \frac{2 + e \cos{\psi}}{1+e \cos{\psi}} \sin{\psi} \mathcal{S}),
\end{align}
\endgroup
where the projections of the perturbation force $f_{\text{perturb}}$ are defined as (these are respectively the component along the separation vector, the one orthogonal to this in the orbital plane and the one along the orbital angular momentum):
\begin{equation}
	\mathcal{S} = \vec{f}_{\text{perturb}} \cdot \frac{\vec{r}}{r}, \qquad \mathcal{W} = \vec{f}_{\text{perturb}} \cdot (\frac{\vec{h}}{h} \times  \frac{\vec{r}}{r}), \qquad \mathcal{R} = \vec{f}_{\text{perturb}} \cdot \frac{\vec{h}}{h}.
\end{equation}

If we simply integrate these evolution equations ``as is'', the solutions will have terms $\sim \psi$ (i.e. not $\sin\psi$ or $\cos\psi$) which indicate a slow, secular evolution over time-scales larger than the orbit. To take these into account more consistently and systematically, it is useful to introduce a multi-scale evolution. The five evolution equations (\ref{eq:pevol})-(\ref{eq:omegaevol}) can be written schematically as:
\be \frac{d\mu^a}{d\psi} = \epsilon\, F^a(\mu_0^b,\psi),\ee
with $F^a$ is the right of the evolution equations and $\mu^a=(p,e,\lambda,\tilde\gamma,\alpha)$ are the five orbit elements excluding $\psi$; the parameter $\epsilon$ is the small parameter that governs the perturbing force.

We then introduce a ``slow scale'' $\tilde\psi := \epsilon\, \psi$, so that the ``total'' $\psi$ derivative becomes:
\be \frac{d}{d\psi} = \frac{\partial}{\partial\psi} + \epsilon \frac{\partial}{\partial\tilde\psi},\ee
and we can then expand the orbital elements to first order as:
\be \label{eq:muaOeps} \mu^a = \mu^a_0(\tilde\psi) + \epsilon\left (\mu_{1,osc}(\psi,\tilde\psi) + \mu_{1,sec}(\tilde\psi)\right) + O(\epsilon^2).\ee
Note that $F=F(\mu_0^a,\psi)$ only depends on the slow scale $\tilde\psi$ through $\mu^a$.
Using that $\partial\mu^a_0/\partial\psi = 0$, $\mu_0^a$ and $\mu_{1,osc}^a$ are then determined by (see \cite{poisson2014gravity} eqs. (12.235) and (12.237)):
\begin{align}
 \frac{\partial\mu_0^a}{\partial \tilde\psi} &= \frac{1}{2\pi} \int_0^{2\pi}d\psi\, F^a(\mu^b_0,\psi),\\
 \mu_{1,osc}^a &= \int d\psi\, \left[ F^a(\mu_0^b,\psi) - \frac{\partial \mu_0^a}{\partial \tilde\psi} \right].
\end{align}
The expansion has been constructed in such a way that $\mu_{1,osc}$ is precisely the $\mathcal{O}(\epsilon)$ piece that is periodically oscillating in $\psi$. At $\mathcal{O}(\epsilon)$ the secular part $\mu_{1,sec}$ is an ``integration constant'' from the perspective of the integral over $\psi$ that determines $\mu_{1,osc}$, and can be chosen freely. (At higher order, it will again be fully determined by $F$, see \cite{poisson2014gravity} eq. (12.238)). We will simply choose $\mu^a_{1,sec} = 0$. This means the secular (not fast-oscillating) part of the orbital element is given to $\mathcal{O}(\epsilon)$ entirely by the zeroth order elements, $\mu^a\approx \mu^a_0$, so that $\mu^a_0$ is the obvious and natural choice to parametrize the orbit evolution with --- in other words, the $e,\lambda,\tilde\gamma,\alpha$ appearing in the evolution equations (\ref{eq:evnu})-(\ref{eq:eve}) are precisely these quantities $\mu_0^a$. If we were to make a different choice for $\mu^a_{1,sec}$, the parametrization would look different in terms of $\mu^a_0$, but would remain identical when expressed in terms of $\mu^a$.

We will also need the period of the orbit, which to $\mathcal{O}(\epsilon)$ is given by:
\be \label{eq:TOeps} T = T_0 + \epsilon\, T_1 = \int_0^{2\pi}d\psi\, \left( \frac{dt}{d\psi}\right)(\mu^a,\psi),\ee
where we need to integrate using the entire  expression (\ref{eq:muaOeps}).

We are interested in the averaged change of the orbital elements over timescales of multiple periods. Then, the $\tilde\psi$ dependence of $\mu_0^a$ is the only important contribution to $\mathcal{O}(\epsilon)$. (Since $\mu_{1,osc}^a$ is periodic in $\psi$, it does not contribute to the averaged, secular evolution.) The averaged, secular change in time of an element is then given to $\mathcal{O}(\epsilon)$ by:
\be \left( \frac{d\mu^a}{dt}\right)_{\text{ave}} = \epsilon\, \frac{2\pi}{T_0}  \frac{\partial\mu^a_0}{\partial \tilde\psi},\ee
where $T_0$ is the $\mathcal{O}(\epsilon^0)$ period in (\ref{eq:TOeps}). This directly gives the conservative, secular ``precession'' effects in $d\mu^a/dt$ in (\ref{eq:evgamma}), (\ref{eq:evalpha}), (\ref{eq:eve}), and (\ref{eq:evlambda}) for the multipolar deformations given by the Lagrangian deformation:
\be \delta \mathcal{L} = -2\mu S_1 \frac{\sin\theta^2\dot{\phi}}{r}  +\mu M_2 \frac{P_2(\cos\theta)}{r^3} -6\mu S_2 \frac{\cos\theta\sin\theta^2\dot{\phi}}{r^2} + \mu M_3 \frac{P_3(\cos\theta)}{r^4},\ee
where each of the $S_i,M_i$ are taken to be $\mathcal{O}(\epsilon)$, and we are using spherical coordinates where the spin $S_1$ is oriented along the $z$-axis. Note that the full Lagrangian is then:
\be \mathcal{L} = \frac{\mu}{2}\dot{\vec{r}}\cdot \dot{\vec{r}} + \frac{\mu M}{r}   + \delta\mathcal{L}.\ee

To calculate the dissipative terms in (\ref{eq:evnu}) for $d\nu/dt$, we use the expression:
\be \label{eq:dEdt} \left(\frac{dE}{dt}\right)_{\text{ave}} =-\frac15 \frac{1}{T} \int_0^{2\pi} d\psi\, \left( \frac{dt}{d\psi}\right) \dddot{M}_{ij}(\mu^a,\psi) \dddot{M}_{ij}(\mu^a,\psi),\ee
to calculate the average rate of energy loss due to radiation over an orbital period, with $T$ the entire expression (\ref{eq:TOeps}). The system quadrupole $M_{ij}(\mu^a,\psi)$ is given by:
\be M_{ij} = [r_ir_j]^{\text{STF}} ,\ee
where STF means we take the symmetric, traceless part. To calculate time derivatives $\dddot{M}_{ij}$, we can use the equations of motion to eliminate any second-order or third-order derivatives, and use (\ref{eq:xv}) to express the remaining zeroth and first derivatives of $x_i(\mu^a,\psi)$ in terms of the elements $\mu^a$. The result (\ref{eq:dEdt}) is a function of the orbital elements $\mu^a$; we then use 
\be \nu(\mu^a) := \frac{1}{T},\ee
to invert the relation and find $p(\nu)$. Finally, we relate $dE/dt$ to $d\nu/dt$ by using:
\be \frac{d\nu}{dt} = \left(\frac{dE}{d\nu}\right)^{-1} \frac{dE}{dt},\ee
where we take $E = (\partial\mathcal{L}/\partial \dot{\vec{r}})\cdot \dot{\vec{r}} -\mathcal{L}$ to be the orbit energy. This then gives the dissipative terms for $M_2,S_2,M_3$ for $d\nu/dt$ in (\ref{eq:evnu}). For the $S_1$ term, we must also take into account the current quadrupole radiation as this contributes at the same order as the mass quadrupole radiation. The only change in procedure is to replace (\ref{eq:dEdt}) by:
\begin{align} \label{eq:dEdtS1}  \left(\frac{dE}{dt}\right)_{\text{ave}} &=-\frac15 \frac{1}{T} \int_0^{2\pi} d\psi\, \left( \frac{dt}{d\psi}\right) \dddot{M}_{ij}(\mu^a,\psi) \dddot{M}_{ij}(\mu^a,\psi)\\
 \nonumber &  -\frac{16}{45} \frac{1}{T} \int_0^{2\pi} d\psi\, \left( \frac{dt}{d\psi}\right) \dddot{S}_{ij}(\mu^a,\psi) \dddot{S}_{ij}(\mu^a,\psi) ,\\
 S_{ij} &= \left[r_i \epsilon_{jkl}r_k\dot{r}_l- \frac32 S_1 r_i \delta_{j3}   \right]^{\text{STF}},
\end{align}
where we are explicitly using that the orientation of the spin $S_1$ is along the $z$-axis in these coordinates.

In practice, to compute the integral in (\ref{eq:dEdt}) or (\ref{eq:dEdtS1}) analytically, we expand and integrate the integrand order by order in $e$. In this way, we find the full expansion in $e$ of the integral after repeating this procedure to a sufficiently high order of $e$ such that the series expansion stops. The exception is one of the contributions of $M_3$, which we only obtain to $\mathcal{O}(e^{16})$ --- see (\ref{eq:M3finiteeguy}).

\bibliographystyle{toine}
\bibliography{oddparity}

\providecommand{\href}[2]{#2}\begingroup\raggedright\begin{thebibliography}{10}

\bibitem{berry2019astro2020}
C.~P. Berry, S.~A. Hughes, C.~F. Sopuerta, A.~J. Chua, A.~Heffernan,
  K.~Holley-Bockelmann, D.~P. Mihaylov, M.~C. Miller  and A.~Sesana10,
  \emph{Astro2020 Science White Paper: The unique potential of extreme
  mass-ratio inspirals for gravitational-wave astronomy}, White paper submitted
  to Astro2020 (2020 Decadal Survey on Astronomy and Astrophysics)
(2019)

\bibitem{baker2019laser}
J.~Baker, J.~Bellovary, P.~L. Bender, E.~Berti, R.~Caldwell, J.~Camp, J.~W.
  Conklin, N.~Cornish, C.~Cutler, R.~DeRosa  {\em et al.}, \emph{The Laser
  Interferometer Space Antenna: unveiling the millihertz gravitational wave
  sky}, arXiv preprint arXiv:1907.06482
(2019)

\bibitem{Barack:2003fp}
L.~Barack and C.~Cutler, \emph{{LISA capture sources: Approximate waveforms,
  signal-to-noise ratios, and parameter estimation accuracy}}, Phys. Rev. D
  {\bf 69} (2004) 082005,
\href{http://www.arXiv.org/abs/gr-qc/0310125}{{\tt gr-qc/0310125}}

\bibitem{Babak:2017tow}
S.~Babak, J.~Gair, A.~Sesana, E.~Barausse, C.~F. Sopuerta, C.~P.~L. Berry,
  E.~Berti, P.~Amaro-Seoane, A.~Petiteau  and A.~Klein, \emph{{Science with the
  space-based interferometer LISA. V: Extreme mass-ratio inspirals}}, Phys.
  Rev. D {\bf 95} (2017), no.~10, 103012,
\href{http://www.arXiv.org/abs/1703.09722}{{\tt 1703.09722}}

\bibitem{Gair:2017ynp}
J.~R. Gair, S.~Babak, A.~Sesana, P.~Amaro-Seoane, E.~Barausse, C.~P.~L. Berry,
  E.~Berti  and C.~Sopuerta, \emph{{Prospects for observing extreme-mass-ratio
  inspirals with LISA}}, J. Phys. Conf. Ser. {\bf 840} (2017), no.~1, 012021,
\href{http://www.arXiv.org/abs/1704.00009}{{\tt 1704.00009}}

\bibitem{Barack:2006pq}
L.~Barack and C.~Cutler, \emph{{Using LISA EMRI sources to test off-Kerr
  deviations in the geometry of massive black holes}}, Phys. Rev. D {\bf 75}
  (2007) 042003,
\href{http://www.arXiv.org/abs/gr-qc/0612029}{{\tt gr-qc/0612029}}

\bibitem{Townsend:1997ku}
P.~K. Townsend, \emph{{Black holes: Lecture notes}},
\href{http://www.arXiv.org/abs/gr-qc/9707012}{{\tt gr-qc/9707012}}

\bibitem{Endlich:2017tqa}
S.~Endlich, V.~Gorbenko, J.~Huang  and L.~Senatore, \emph{{An effective
  formalism for testing extensions to General Relativity with gravitational
  waves}}, JHEP {\bf 09} (2017) 122,
\href{http://www.arXiv.org/abs/1704.01590}{{\tt 1704.01590}}

\bibitem{Cardoso:2018ptl}
V.~Cardoso, M.~Kimura, A.~Maselli  and L.~Senatore, \emph{{Black Holes in an
  Effective Field Theory Extension of General Relativity}}, Phys. Rev. Lett.
  {\bf 121} (2018), no.~25, 251105,
\href{http://www.arXiv.org/abs/1808.08962}{{\tt 1808.08962}}

\bibitem{Cano:2019ore}
P.~A. Cano and A.~Ruip\'erez, \emph{{Leading higher-derivative corrections to
  Kerr geometry}}, JHEP {\bf 05} (2019) 189,
  \href{http://www.arXiv.org/abs/1901.01315}{{\tt 1901.01315}},
[Erratum: JHEP 03, 187 (2020)]

\bibitem{Bena:2020uup}
I.~Bena and D.~R. Mayerson, \emph{{Black Holes Lessons from Multipole Ratios}},
  JHEP {\bf 03} (2021) 114,
\href{http://www.arXiv.org/abs/2007.09152}{{\tt 2007.09152}}

\bibitem{Bah:2021jno}
I.~Bah, I.~Bena, P.~Heidmann, Y.~Li  and D.~R. Mayerson, \emph{{Gravitational
  footprints of black holes and their microstate geometries}}, JHEP {\bf 10}
  (2021) 138,
\href{http://www.arXiv.org/abs/2104.10686}{{\tt 2104.10686}}

\bibitem{Bena:2020see}
I.~Bena and D.~R. Mayerson, \emph{{Multipole Ratios: A New Window into Black
  Holes}}, Phys. Rev. Lett. {\bf 125} (2020), no.~22, 221602,
\href{http://www.arXiv.org/abs/2006.10750}{{\tt 2006.10750}}

\bibitem{Bianchi:2020bxa}
M.~Bianchi, D.~Consoli, A.~Grillo, J.~F. Morales, P.~Pani  and G.~Raposo,
  \emph{{Distinguishing fuzzballs from black holes through their multipolar
  structure}}, Phys. Rev. Lett. {\bf 125} (2020), no.~22, 221601,
\href{http://www.arXiv.org/abs/2007.01743}{{\tt 2007.01743}}

\bibitem{Bianchi:2020miz}
M.~Bianchi, D.~Consoli, A.~Grillo, J.~F. Morales, P.~Pani  and G.~Raposo,
  \emph{{The multipolar structure of fuzzballs}}, JHEP {\bf 01} (2021) 003,
\href{http://www.arXiv.org/abs/2008.01445}{{\tt 2008.01445}}

\bibitem{Mayerson:2020tpn}
D.~R. Mayerson, \emph{{Fuzzballs and Observations}}, Gen. Rel. Grav. {\bf 52}
  (2020), no.~12, 115,
\href{http://www.arXiv.org/abs/2010.09736}{{\tt 2010.09736}}

\bibitem{Ryan:1995wh}
F.~D. Ryan, \emph{{Gravitational waves from the inspiral of a compact object
  into a massive, axisymmetric body with arbitrary multipole moments}}, Phys.
  Rev. D {\bf 52} (1995)
5707--5718

\bibitem{Glampedakis:2005cf}
K.~Glampedakis and S.~Babak, \emph{{Mapping spacetimes with LISA: Inspiral of a
  test-body in a `quasi-Kerr' field}}, Class. Quant. Grav. {\bf 23} (2006)
  4167--4188,
\href{http://www.arXiv.org/abs/gr-qc/0510057}{{\tt gr-qc/0510057}}

\bibitem{Gair:2007kr}
J.~R. Gair, C.~Li  and I.~Mandel, \emph{{Observable Properties of Orbits in
  Exact Bumpy Spacetimes}}, Phys. Rev. D {\bf 77} (2008) 024035,
\href{http://www.arXiv.org/abs/0708.0628}{{\tt 0708.0628}}

\bibitem{Moore:2017lxy}
C.~J. Moore, A.~J.~K. Chua  and J.~R. Gair, \emph{{Gravitational waves from
  extreme mass ratio inspirals around bumpy black holes}}, Class. Quant. Grav.
  {\bf 34} (2017), no.~19, 195009,
\href{http://www.arXiv.org/abs/1707.00712}{{\tt 1707.00712}}

\bibitem{Raposo:2018xkf}
G.~Raposo, P.~Pani  and R.~Emparan, \emph{{Exotic compact objects with soft
  hair}}, Phys. Rev. D {\bf 99} (2019), no.~10, 104050,
\href{http://www.arXiv.org/abs/1812.07615}{{\tt 1812.07615}}

\bibitem{Cunha:2018uzc}
P.~V.~P. Cunha, C.~A.~R. Herdeiro  and E.~Radu, \emph{{Isolated black holes
  without $\mathbb Z_2$ isometry}}, Phys. Rev. D {\bf 98} (2018), no.~10,
  104060,
\href{http://www.arXiv.org/abs/1808.06692}{{\tt 1808.06692}}

\bibitem{Datta:2020axm}
S.~Datta and S.~Mukherjee, \emph{{Possible connection between the reflection
  symmetry and existence of equatorial circular orbit}}, Phys. Rev. D {\bf 103}
  (2021), no.~10, 104032,
\href{http://www.arXiv.org/abs/2010.12387}{{\tt 2010.12387}}

\bibitem{Lima:2021las}
H.~C.~D. Lima, Junior., L.~C.~B. Crispino, P.~V.~P. Cunha  and C.~A.~R.
  Herdeiro, \emph{{Can different black holes cast the same shadow?}}, Phys.
  Rev. D {\bf 103} (2021), no.~8, 084040,
\href{http://www.arXiv.org/abs/2102.07034}{{\tt 2102.07034}}

\bibitem{Mukherjee:2020how}
S.~Mukherjee and S.~Chakraborty, \emph{{Multipole moments of compact objects
  with NUT charge: Theoretical and observational implications}}, Phys. Rev. D
  {\bf 102} (2020) 124058,
\href{http://www.arXiv.org/abs/2008.06891}{{\tt 2008.06891}}

\bibitem{Sopuerta:2009iy}
C.~F. Sopuerta and N.~Yunes, \emph{{Extreme and Intermediate-Mass Ratio
  Inspirals in Dynamical Chern-Simons Modified Gravity}}, Phys. Rev. D {\bf 80}
  (2009) 064006,
\href{http://www.arXiv.org/abs/0904.4501}{{\tt 0904.4501}}

\bibitem{Vigeland:2010xe}
S.~J. Vigeland, \emph{{Multipole moments of bumpy black holes}}, Phys. Rev. D
  {\bf 82} (2010) 104041,
\href{http://www.arXiv.org/abs/1008.1278}{{\tt 1008.1278}}

\bibitem{Vigeland:2009pr}
S.~J. Vigeland and S.~A. Hughes, \emph{{Spacetime and orbits of bumpy black
  holes}}, Phys. Rev. D {\bf 81} (2010) 024030,
\href{http://www.arXiv.org/abs/0911.1756}{{\tt 0911.1756}}

\bibitem{Ryan:1997hg}
F.~D. Ryan, \emph{{Accuracy of estimating the multipole moments of a massive
  body from the gravitational waves of a binary inspiral}}, Phys. Rev. D {\bf
  56} (1997)
1845--1855

\bibitem{Buonanno:2009zt}
A.~Buonanno, B.~Iyer, E.~Ochsner, Y.~Pan  and B.~S. Sathyaprakash,
  \emph{{Comparison of post-Newtonian templates for compact binary inspiral
  signals in gravitational-wave detectors}}, Phys. Rev. D {\bf 80} (2009)
  084043,
\href{http://www.arXiv.org/abs/0907.0700}{{\tt 0907.0700}}

\bibitem{tagoshi1994post}
H.~Tagoshi and M.~Sasaki, \emph{Post-Newtonian expansion of gravitational waves
  from a particle in circular orbit around a Schwarzschild black hole},
  Progress of Theoretical Physics {\bf 92} (1994), no.~4,
745--771

\bibitem{Poisson:1994yf}
E.~Poisson and M.~Sasaki, \emph{{Gravitational radiation from a particle in
  circular orbit around a black hole. 5: Black hole absorption and tail
  corrections}}, Phys. Rev. D {\bf 51} (1995) 5753--5767,
\href{http://www.arXiv.org/abs/gr-qc/9412027}{{\tt gr-qc/9412027}}

\bibitem{Poisson:1995vs}
E.~Poisson, \emph{{Gravitational radiation from a particle in circular orbit
  around a black hole. 6. Accuracy of the postNewtonian expansion}}, Phys. Rev.
  D {\bf 52} (1995) 5719--5723,
  \href{http://www.arXiv.org/abs/gr-qc/9505030}{{\tt gr-qc/9505030}},
[Addendum: Phys.Rev.D 55, 7980--7981 (1997)]

\bibitem{Tanaka:1996lfd}
T.~Tanaka, H.~Tagoshi  and M.~Sasaki, \emph{{Gravitational waves by a particle
  in circular orbits around a Schwarzschild black hole: 5.5 postNewtonian
  formula}}, Prog. Theor. Phys. {\bf 96} (1996) 1087--1101,
\href{http://www.arXiv.org/abs/gr-qc/9701050}{{\tt gr-qc/9701050}}

\bibitem{Robson:2018ifk}
T.~Robson, N.~J. Cornish  and C.~Liu, \emph{{The construction and use of LISA
  sensitivity curves}}, Class. Quant. Grav. {\bf 36} (2019), no.~10, 105011,
\href{http://www.arXiv.org/abs/1803.01944}{{\tt 1803.01944}}

\bibitem{Poisson:2011nh}
E.~Poisson, A.~Pound  and I.~Vega, \emph{{The Motion of point particles in
  curved spacetime}}, Living Rev. Rel. {\bf 14} (2011) 7,
\href{http://www.arXiv.org/abs/1102.0529}{{\tt 1102.0529}}

\bibitem{Barack:2018yvs}
L.~Barack and A.~Pound, \emph{{Self-force and radiation reaction in general
  relativity}}, Rept. Prog. Phys. {\bf 82} (2019), no.~1, 016904,
\href{http://www.arXiv.org/abs/1805.10385}{{\tt 1805.10385}}

\bibitem{Pound:2021qin}
A.~Pound and B.~Wardell, \emph{{Black hole perturbation theory and
  gravitational self-force}},
\href{http://www.arXiv.org/abs/2101.04592}{{\tt 2101.04592}}

\bibitem{Hughes:2021exa}
S.~A. Hughes, N.~Warburton, G.~Khanna, A.~J.~K. Chua  and M.~L. Katz,
  \emph{{Adiabatic waveforms for extreme mass-ratio inspirals via multivoice
  decomposition in time and frequency}}, Phys. Rev. D {\bf 103} (2021), no.~10,
  104014,
\href{http://www.arXiv.org/abs/2102.02713}{{\tt 2102.02713}}

\bibitem{vandeMeent:2017bcc}
M.~van~de Meent, \emph{{Gravitational self-force on generic bound geodesics in
  Kerr spacetime}}, Phys. Rev. D {\bf 97} (2018), no.~10, 104033,
\href{http://www.arXiv.org/abs/1711.09607}{{\tt 1711.09607}}

\bibitem{Warburton:2021kwk}
N.~Warburton, A.~Pound, B.~Wardell, J.~Miller  and L.~Durkan,
  \emph{{Gravitational-Wave Energy Flux for Compact Binaries through Second
  Order in the Mass Ratio}}, Phys. Rev. Lett. {\bf 127} (2021), no.~15, 151102,
\href{http://www.arXiv.org/abs/2107.01298}{{\tt 2107.01298}}

\bibitem{Wardell:2021fyy}
B.~Wardell, A.~Pound, N.~Warburton, J.~Miller, L.~Durkan  and A.~L. Tiec,
  \emph{{Gravitational waveforms for compact binaries from second-order
  self-force theory}},
\href{http://www.arXiv.org/abs/2112.12265}{{\tt 2112.12265}}

\bibitem{VanDeMeent:2018cgn}
M.~Van De~Meent and N.~Warburton, \emph{{Fast Self-forced Inspirals}}, Class.
  Quant. Grav. {\bf 35} (2018), no.~14, 144003,
\href{http://www.arXiv.org/abs/1802.05281}{{\tt 1802.05281}}

\bibitem{Chua:2020stf}
A.~J.~K. Chua, M.~L. Katz, N.~Warburton  and S.~A. Hughes, \emph{{Rapid
  generation of fully relativistic extreme-mass-ratio-inspiral waveform
  templates for LISA data analysis}}, Phys. Rev. Lett. {\bf 126} (2021), no.~5,
  051102,
\href{http://www.arXiv.org/abs/2008.06071}{{\tt 2008.06071}}

\bibitem{Katz:2021yft}
M.~L. Katz, A.~J.~K. Chua, L.~Speri, N.~Warburton  and S.~A. Hughes,
  \emph{{Fast extreme-mass-ratio-inspiral waveforms: New tools for millihertz
  gravitational-wave data analysis}}, Phys. Rev. D {\bf 104} (2021), no.~6,
  064047,
\href{http://www.arXiv.org/abs/2104.04582}{{\tt 2104.04582}}

\bibitem{Glampedakis:2002cb}
K.~Glampedakis, S.~A. Hughes  and D.~Kennefick, \emph{{Approximating the
  inspiral of test bodies into Kerr black holes}}, Phys. Rev. D {\bf 66} (2002)
  064005,
\href{http://www.arXiv.org/abs/gr-qc/0205033}{{\tt gr-qc/0205033}}

\bibitem{Glampedakis:2002ya}
K.~Glampedakis and D.~Kennefick, \emph{{Zoom and whirl: Eccentric equatorial
  orbits around spinning black holes and their evolution under gravitational
  radiation reaction}}, Phys. Rev. D {\bf 66} (2002) 044002,
\href{http://www.arXiv.org/abs/gr-qc/0203086}{{\tt gr-qc/0203086}}

\bibitem{Gair:2005ih}
J.~R. Gair and K.~Glampedakis, \emph{{Improved approximate inspirals of
  test-bodies into Kerr black holes}}, Phys. Rev. D {\bf 73} (2006) 064037,
\href{http://www.arXiv.org/abs/gr-qc/0510129}{{\tt gr-qc/0510129}}

\bibitem{Babak:2006uv}
S.~Babak, H.~Fang, J.~R. Gair, K.~Glampedakis  and S.~A. Hughes,
  \emph{{'Kludge' gravitational waveforms for a test-body orbiting a Kerr black
  hole}}, Phys. Rev. D {\bf 75} (2007) 024005,
  \href{http://www.arXiv.org/abs/gr-qc/0607007}{{\tt gr-qc/0607007}},
[Erratum: Phys.Rev.D 77, 04990 (2008)]

\bibitem{Chua:2015mua}
A.~J.~K. Chua and J.~R. Gair, \emph{{Improved analytic extreme-mass-ratio
  inspiral model for scoping out eLISA data analysis}}, Class. Quant. Grav.
  {\bf 32} (2015) 232002,
\href{http://www.arXiv.org/abs/1510.06245}{{\tt 1510.06245}}

\bibitem{Chua:2017ujo}
A.~J.~K. Chua, C.~J. Moore  and J.~R. Gair, \emph{{Augmented kludge waveforms
  for detecting extreme-mass-ratio inspirals}}, Phys. Rev. D {\bf 96} (2017),
  no.~4, 044005,
\href{http://www.arXiv.org/abs/1705.04259}{{\tt 1705.04259}}

\bibitem{Liu:2020ghq}
M.~Liu and J.-d. Zhang, \emph{{Augmented analytic kludge waveform with
  quadrupole moment correction}},
\href{http://www.arXiv.org/abs/2008.11396}{{\tt 2008.11396}}

\bibitem{Yunes:2009ef}
N.~Yunes, A.~Buonanno, S.~A. Hughes, M.~Coleman~Miller  and Y.~Pan,
  \emph{{Modeling Extreme Mass Ratio Inspirals within the Effective-One-Body
  Approach}}, Phys. Rev. Lett. {\bf 104} (2010) 091102,
\href{http://www.arXiv.org/abs/0909.4263}{{\tt 0909.4263}}

\bibitem{Yunes:2010zj}
N.~Yunes, A.~Buonanno, S.~A. Hughes, Y.~Pan, E.~Barausse, M.~C. Miller  and
  W.~Throwe, \emph{{Extreme Mass-Ratio Inspirals in the Effective-One-Body
  Approach: Quasi-Circular, Equatorial Orbits around a Spinning Black Hole}},
  Phys. Rev. D {\bf 83} (2011) 044044,
  \href{http://www.arXiv.org/abs/1009.6013}{{\tt 1009.6013}},
[Erratum: Phys.Rev.D 88, 109904 (2013)]

\bibitem{Albanesi:2021rby}
S.~Albanesi, A.~Nagar  and S.~Bernuzzi, \emph{{Effective one-body model for
  extreme-mass-ratio spinning binaries on eccentric equatorial orbits: Testing
  radiation reaction and waveform}}, Phys. Rev. D {\bf 104} (2021), no.~2,
  024067,
\href{http://www.arXiv.org/abs/2104.10559}{{\tt 2104.10559}}

\bibitem{Huerta:2008gb}
E.~A. Huerta and J.~R. Gair, \emph{{Influence of conservative corrections on
  parameter estimation for extreme-mass-ratio inspirals}}, Phys. Rev. D {\bf
  79} (2009) 084021, \href{http://www.arXiv.org/abs/0812.4208}{{\tt
  0812.4208}},
[Erratum: Phys.Rev.D 84, 049903 (2011)]

\bibitem{Cutler:1997ta}
C.~Cutler, \emph{{Angular resolution of the LISA gravitational wave detector}},
  Phys. Rev. D {\bf 57} (1998) 7089--7102,
\href{http://www.arXiv.org/abs/gr-qc/9703068}{{\tt gr-qc/9703068}}

\bibitem{goldstein}
H.~Goldstein, C.~Poole  and J.~Safko, {\em Classical Mechanics}.
\newblock Addison-Wesley, Boston, 3rd~ed.,
2002

\bibitem{poisson2014gravity}
E.~Poisson and C.~M. Will, {\em Gravity: Newtonian, post-newtonian,
  relativistic}.
\newblock Cambridge University Press,
2014

\bibitem{Hughes:1999bq}
S.~A. Hughes, \emph{{The Evolution of circular, nonequatorial orbits of Kerr
  black holes due to gravitational wave emission}}, Phys. Rev. D {\bf 61}
  (2000), no.~8, 084004, \href{http://www.arXiv.org/abs/gr-qc/9910091}{{\tt
  gr-qc/9910091}},
[Erratum: Phys.Rev.D 63, 049902 (2001), Erratum: Phys.Rev.D 65, 069902 (2002),
  Erratum: Phys.Rev.D 67, 089901 (2003), Erratum: Phys.Rev.D 78, 109902 (2008),
  Erratum: Phys.Rev.D 90, 109904 (2014)]

\bibitem{Ryan:1995xi}
F.~D. Ryan, \emph{{Effect of gravitational radiation reaction on nonequatorial
  orbits around a Kerr black hole}}, Phys. Rev. D {\bf 53} (1996) 3064--3069,
\href{http://www.arXiv.org/abs/gr-qc/9511062}{{\tt gr-qc/9511062}}

\bibitem{peters1963gravitational}
P.~C. Peters and J.~Mathews, \emph{Gravitational radiation from point masses in
  a Keplerian orbit}, Physical Review {\bf 131} (1963), no.~1,
435

\bibitem{peters1964gravitational}
P.~C. Peters, \emph{Gravitational radiation and the motion of two point
  masses}, Physical Review {\bf 136} (1964), no.~4B,
B1224

\bibitem{BHPToolkit}
\emph{{Black Hole Perturbation Toolkit}t.}
  (\href{http://bhptoolkit.org/}{bhptoolkit.org}).

\bibitem{Gair:2004iv}
J.~R. Gair, L.~Barack, T.~Creighton, C.~Cutler, S.~L. Larson, E.~S. Phinney
  and M.~Vallisneri, \emph{{Event rate estimates for LISA extreme mass ratio
  capture sources}}, Class. Quant. Grav. {\bf 21} (2004) S1595--S1606,
\href{http://www.arXiv.org/abs/gr-qc/0405137}{{\tt gr-qc/0405137}}

\bibitem{babak2010mock}
S.~Babak, J.~G. Baker, M.~J. Benacquista, N.~J. Cornish, S.~L. Larson,
  I.~Mandel, S.~T. McWilliams, A.~Petiteau, E.~K. Porter, E.~L. Robinson  {\em
  et al.}, \emph{The Mock LISA Data Challenges: from challenge 3 to challenge
  4}, Classical and Quantum Gravity {\bf 27} (2010), no.~8,
084009

\bibitem{upcomingHDmultipoles}
P.~Cano, B.~Ganchev, D.~R. Mayerson  and A.~Ruip\'erez~Vicente.
\emph{In preparation}

\bibitem{Ryan:1995zm}
F.~D. Ryan, \emph{{Effect of gravitational radiation reaction on circular
  orbits around a spinning black hole}}, Phys. Rev. D {\bf 52} (1995)
  R3159--R3162,
\href{http://www.arXiv.org/abs/gr-qc/9506023}{{\tt gr-qc/9506023}}

\bibitem{ernst1968new}
F.~J. Ernst, \emph{New formulation of the axially symmetric gravitational field
  problem}, Physical Review {\bf 167} (1968), no.~5,
1175

\bibitem{meinel1995asymptotically}
R.~Meinel and G.~Neugebauer, \emph{Asymptotically flat solutions to the Ernst
  equation with reflection symmetry}, Classical and Quantum Gravity {\bf 12}
  (1995), no.~8,
2045

\bibitem{kordas1995reflection}
P.~Kordas, \emph{Reflection-symmetric, asymptotically flat solutions of the
  vacuum axistationary Einstein equations}, Classical and Quantum Gravity {\bf
  12} (1995), no.~8,
2037

\bibitem{fodor1989multipole}
G.~Fodor, C.~Hoenselaers  and Z.~Perj{\'e}s, \emph{Multipole moments of
  axisymmetric systems in relativity}, Journal of Mathematical Physics {\bf 30}
  (1989), no.~10,
2252--2257

\bibitem{sotiriou2004corrections}
T.~P. Sotiriou and T.~A. Apostolatos, \emph{Corrections and comments on the
  multipole moments of axisymmetric electrovacuum spacetimes}, Classical and
  Quantum Gravity {\bf 21} (2004), no.~24,
5727

\bibitem{fodor2021calculation}
G.~Fodor, E.~dos Santos Costa~Filho  and B.~Hartmann, \emph{Calculation of
  multipole moments of axistationary electrovacuum spacetimes}, Physical Review
  D {\bf 104} (2021), no.~6,
064012

\bibitem{geroch1970multipole}
R.~Geroch, \emph{Multipole moments. II. Curved space}, Journal of Mathematical
  Physics {\bf 11} (1970), no.~8,
2580--2588

\bibitem{hansen1974multipole}
R.~O. Hansen, \emph{Multipole moments of stationary space-times}, Journal of
  Mathematical Physics {\bf 15} (1974), no.~1,
46--52

\bibitem{Cornish:2017vip}
N.~Cornish and T.~Robson, \emph{{Galactic binary science with the new LISA
  design}}, J. Phys. Conf. Ser. {\bf 840} (2017), no.~1, 012024,
\href{http://www.arXiv.org/abs/1703.09858}{{\tt 1703.09858}}

\bibitem{Junker:1992kle}
W.~Junker and G.~Sch\"afer, \emph{{Binary systems: higher order gravitational
  radiation damping and wave emission}}, Mon. Not. Roy. Astron. Soc. {\bf 254}
  (1992), no.~1,
146--164

\bibitem{Lai:1995hi}
D.~Lai, L.~Bildsten  and V.~M. Kaspi, \emph{{Spin orbit interaction in neutron
  star / main sequence binaries and implications for pulsar timing}},
  Astrophys. J. {\bf 452} (1995) 819,
\href{http://www.arXiv.org/abs/astro-ph/9505042}{{\tt astro-ph/9505042}}

\end{thebibliography}\endgroup

\end{document}